\documentclass[%
 aip,
 amsmath,amssymb,
 reprint,%
]{revtex4-1}
\usepackage{graphicx}% Include figure files
\usepackage{subfigure}
\usepackage{dcolumn}% Align table columns on decimal point
\usepackage{bm}% bold math
\usepackage{color}
\usepackage[T1]{fontenc}
\usepackage{mathptmx}
\usepackage{multirow}
\usepackage{array}
\usepackage{float}
\usepackage{etoolbox}
\usepackage{booktabs}
\usepackage{url}
\usepackage{CJKutf8}
\usepackage{pifont}
\usepackage{epstopdf, epsfig}
\usepackage{graphicx}
\usepackage{longtable}
\usepackage[
    colorlinks=true,
    linkcolor=blue,
    urlcolor=blue,
    citecolor=blue
]{hyperref}

\usepackage{titlesec}
\titleclass{\subsubsubsection}{straight}[\subsection]
\newcounter{subsubsubsection}[subsubsection]
\renewcommand\thesubsubsubsection{\thesubsubsection.\alph{subsubsubsection}}
\titleformat{\subsubsubsection}
 {\normalfont\normalsize\bfseries}{\thesubsubsubsection}{1em}{}
\titlespacing*{\subsubsubsection}
{0pt}{3.25ex plus 1ex minus .2ex}{1.5ex plus .2ex}

\begin{document}
%\begin{CJK*}{UTF8}{gbsn}
\preprint{AIP/123-QED}

%\title{Multiscale modeling and simulation of thermodynamic nonequilibrium effects in three-dimensional high-speed compressible flows: Based on the discrete Boltzmann method}

\title{Thermodynamic nonequilibrium effects in three-dimensional high-speed compressible flows: Multiscale modeling and simulation via the discrete Boltzmann method}

\author{Qinghong Guo \begin{CJK*}{UTF8}{gbsn} (郭清红) \end{CJK*}}
 \affiliation{School of Mathematics and Statistics, Key Laboratory of Analytical Mathematics and Applications (Ministry of Education), Fujian Key Laboratory of Analytical Mathematics and Applications (FJKLAMA), Center for Applied Mathematics of Fujian Province (FJNU), Fujian Normal University, 350117 Fuzhou, China}
\author{Yanbiao Gan \begin{CJK*}{UTF8}{gbsn} (甘延标) \end{CJK*}}
 \thanks{Corresponding author: gan@nciae.edu.cn}
  \affiliation{Hebei Key Laboratory of Trans-Media Aerial Underwater Vehicle,
North China Institute of Aerospace Engineering, Langfang 065000, China}
\author{Bin Yang \begin{CJK*}{UTF8}{gbsn} (杨斌) \end{CJK*}}
  \affiliation{School of Energy and Safety Engineering, Tianjin Chengjian University, Tianjin 300384, China}
\author{Yanhong Wu \begin{CJK*}{UTF8}{gbsn} (吴彦宏) \end{CJK*}}
  \affiliation{Hebei Key Laboratory of Trans-Media Aerial Underwater Vehicle,
North China Institute of Aerospace Engineering, Langfang 065000, China}
\author{Huilin Lai \begin{CJK*}{UTF8}{gbsn} (赖惠林)\end{CJK*}}
 \thanks{Corresponding author: hllai@fjnu.edu.cn}
\affiliation{School of Mathematics and Statistics, Key Laboratory of Analytical Mathematics and Applications (Ministry of Education), Fujian Key Laboratory of Analytical Mathematics and Applications (FJKLAMA), Center for Applied Mathematics of Fujian Province (FJNU), Fujian Normal University, 350117 Fuzhou, China}
\author{Aiguo Xu \begin{CJK*}{UTF8}{gbsn} (许爱国) \end{CJK*}}
\affiliation{National Key Laboratory of Computational Physics, Institute of Applied Physics and Computational Mathematics, P. O. Box 8009-26, Beijing 100088, P.R.China}
\affiliation{National Key Laboratory of Shock Wave and Detonation Physics, Mianyang 621999, China}
\affiliation{HEDPS, Center for Applied Physics and Technology, and College of Engineering, Peking University, Beijing 100871, China}
\affiliation{State Key Laboratory of Explosion Science and Safety Protection, Beijing Institute of Technology, Beijing 100081, China}

\date{\today}% It is always \today, today,

\begin{abstract}
Three-dimensional (3D) high-speed compressible flow is a typical nonlinear, nonequilibrium, and multiscale complex flow. Traditional fluid mechanics models, based on the quasi-continuum assumption and near-equilibrium approximation, are insufficient to capture significant discrete effects and thermodynamic nonequilibrium effects (TNEs) as the Knudsen number increases.
To overcome these limitations, a discrete Boltzmann modeling and simulation method, rooted in  kinetic and mean-field theories, has been developed.
By applying Chapman-Enskog multiscale analysis, the essential kinetic moment relations $\bm{\Phi}=(\mathbf{M}_0,\mathbf{M}_1,\mathbf{M}_{2,0},\mathbf{M}_2,\mathbf{M}_{3,1},\mathbf{M}_3,
\mathbf{M}_{4,2},\mathbf{M}_4,\mathbf{M}_{5,3})$ for characterizing second-order TNEs are determined.
These relations $\bm{\Phi}$ are invariants in coarse-grained physical modeling, providing a unique mesoscopic perspective for analyzing TNE behaviors.
A discrete Boltzmann model, accurate to second-order in the Knudsen number, is developed to enable multiscale simulations of 3D supersonic flows.
As key TNE measures, nonlinear constitutive relations (NCRs), are theoretically derived for the 3D case, offering a constitutive foundation for improving macroscopic fluid modeling.
The NCRs in three dimensions exhibit greater complexity than their two-dimensional counterparts.
This complexity arises from increased degrees of freedom, which introduce additional kinds of nonequilibrium driving forces, stronger coupling between these forces, and a significant increase in nonequilibrium components.
At the macroscopic level, the model is validated through a series of classical test cases, ranging from one-dimensional to 3D scenarios, from subsonic to supersonic regimes.
At the mesoscopic level, the model accurately captures typical TNEs, such as viscous stress and heat flux, around mesoscale structures, across various scales and orders. This work provides kinetic insights that advance  multiscale simulation techniques for 3D high-speed compressible flows.
\end{abstract}

\maketitle
%\end{CJK*}

\section{\label{sec:level1} Introduction}

High-speed compressible flow arises when fluid velocity approaches or exceeds the speed of sound, leading to significant density variations. Such flows exhibit strong nonlinearity, nonequilibrium effects, and multiscale characteristics, appearing in aerospace, defense engineering, natural phenomena, and even daily life.

Understanding and controlling high-speed compressible flow is of great scientific and engineering significance\cite{sasoh2020CFDSW}.
In aerospace applications, hypersonic vehicles typically operate at speeds exceeding Mach 5, generating shock waves, rarefaction waves, boundary layers, and turbulent regions around surfaces and within intake ducts, with pronounced nonequilibrium effects at the mesoscopic scale.
Shock waves induce sudden pressure rises and intense thermal loads, which can compromise structural integrity and lead to vehicle failure.
Moreover, shock-wave–boundary-layer interactions often lead to separation, a phenomenon that increases drag and reduces aerodynamic stability.
A thorough understanding of high-speed compressible flow mechanisms is crucial for optimizing vehicle design, minimizing aerodynamic drag and thermal loads, and enhancing stability and maneuverability\cite{anderson1989Aiaa,Zhang2020PAS,peters2024NC}.

During the implosion acceleration stage of inertial confinement fusion, the implosion velocity can reach $300$–$400$ km/s. In this process, isotropic shock wave propagation and multiscale coupling in high-speed compressible flows play a pivotal role in achieving efficient energy compression\cite{mostert2015POF,macphee2018POP,roycroft2022POP}.
In defense engineering, detonation and shock phenomena are fundamental to a wide range of applications. Detonation serves as a cornerstone for the design and development of conventional and advanced weaponry, supporting technologies such as armor penetration, blast protection, engineering blasting, and explosive processing.
The expansion velocities of detonation products can range from thousands to tens of thousands of meters per second, with pressures reaching several hundred thousand atmospheres, making these phenomena quintessential examples of high-speed compressible flow.
Research on shock and detonation phenomena increasingly focuses on fine physical structures and nonequilibrium behaviors near interfaces, including shock waves, detonation waves, and rarefaction waves\cite{Lee2008CUP,Rosato2021NAS}. These dynamic behaviors highlight the importance of understanding nonequilibrium mechanisms in extreme high-speed flow environments.

Beyond applications in aerospace, energy technology, and weapon systems, exploring high-speed compressible flow mechanisms helps overcome the limitations of traditional hydrodynamic models in small-scale, high-pressure conditions.
For instance, in high-performance chip cooling within micro-electromechanical systems, high-speed gas flowing through micro-nozzles generates localized jets on the chip surface. These jets enhance convective heat transfer by disrupting the thermal boundary layer through shear forces. Additionally, compressible effects at the nozzle exit generate expansion waves, reducing gas temperature, improving cooling efficiency, and mitigating shock wave effects\cite{ho1998AROFM,stone2004ARFM}.
In heap leaching with microporous seepage, high-pressure fluids moving through micropores and fractures induce local pressure and velocity fluctuations. These fluctuations enhance pore wall dissolution and facilitate fracture propagation. Shock waves at pore tips further assist in forming interconnected fracture pathways\cite{Wang2018JFM}.
These examples highlight the critical role of high-speed compressible flow in small-scale nonequilibrium processes, where the interplay of high pressures, localized effects, and compressibility drives significant physical and engineering effects.

However, research on high-speed compressible flow faces numerous complex challenges, including: (i) intense hydrodynamic and thermodynamic nonequilibrium effects (HNEs and TNEs, respectively); (ii) spatiotemporal multiscale coupling, from microscopic molecular collisions to macroscopic turbulent structures; (iii) pronounced nonlinear flow effects; (iv) highly complex boundary conditions. Therefore, a key challenge is developing multiscale models that accurately describe TNEs across different scales.

The Knudsen number ($Kn$) is a key dimensionless parameter that quantifies the nonequilibrium characteristics of fluid flows. It is defined as the ratio of the molecular mean free path to the characteristic flow length. Based on $Kn$, flow regimes are typically divided into four categories: (i) continuum flow ($Kn < 0.001$), (ii) slip flow ($0.001 < Kn < 0.1$), (iii) transitional flow ($0.1 < Kn < 10$), and (iv) free molecular flow ($Kn > 10$).

In the continuum flow regime, the Navier-Stokes (NS) equations, formulated under the continuum assumption and near-equilibrium approximation, serve as the primary analytical tools.
These equations, incorporating linear constitutive relations (e.g., Newton's viscosity law, Fourier's heat conduction law), remain valid only under weak TNE conditions.
In the slip flow regime, velocity slip and temperature jump at boundaries necessitate specific slip corrections to the boundary conditions\cite{smoluchowski1898ADP}.
As flows enter the transitional and free molecular regimes, the limitations of the NS equations become apparent due to their inability to adequately describe strong nonequilibrium effects. In these regimes, kinetic theory-based methods are essential for accurately modeling multiscale flow behavior.

The Liouville equation is fundamental to nonequilibrium statistical physics and serves as the foundation for mesoscopic kinetic modeling. The Boltzmann equation, derived from a coarse-grained Liouville equation, is the most widely used equation in kinetic modeling. As a cornerstone of nonequilibrium statistical physics, the Boltzmann equation links microscopic particle behavior to macroscopic flow dynamics, making it particularly effective for modeling rarefied gases and strongly nonequilibrium flows\cite{Xu2022}.
However, the Boltzmann equation is a seven-dimensional integro-differential equation involving time, space, velocity, and complex collision terms, making its numerical solution extremely challenging.  In the transitional flow regime, where molecular collisions are infrequent, Bird’s direct simulation Monte Carlo (DSMC) method is widely used to solve the Boltzmann equation\cite{bird1994OUP,shen2006SSBM,virgile2022AA}.
Besides particle-based probabilistic methods like DSMC, deterministic numerical approaches have been developed for directly solving the Boltzmann equation. These methods include the fast spectral method\cite{mouhot2006MOC,Wu2015POF} and the Fourier spectral method\cite{bobylev1988SSRSCMPR,Wu2017CAS}. Although these methods accurately capture phenomena such as thermal fluctuations in microscopic particles, their high computational cost restricts them to small-scale problems. Scaling these methods to larger spatial and temporal domains remains challenging.

Due to the complexity of the Boltzmann equation's collision term, developing simplified kinetic models that capture its most relevant features is an effective approach. Common simplified models include the Bhatnagar-Gross-Krook (BGK) model\cite{bhatnagar1954PR}, the ellipsoidal statistical BGK (ES-BGK) model\cite{holway1966TPOF}, the Shakhov model\cite{shakhov1968FD}, and the Liu model\cite{Liu1990POF}.
To further enhance computational efficiency and applicability, researchers have explored various strategies for simplifying collision terms and tailoring models to specific problems. This has resulted in the development of various computational methods for solving cross-scale flow problems. Notable examples include the gas-kinetic scheme\cite{Xu1994JOCP,Xu2001JOCP}, unified gas-kinetic scheme\cite{Xu2010JOCP,Huang2012CICP}, discrete unified gas-kinetic scheme\cite{Guo2013PRE,Xin2023AIA}, discrete velocity method\cite{Yang2016JOCP,Yuan2020CPC,yang2022pof}, and the lattice Boltzmann method (LBM)\cite{He1997PRE,Gan2008PA,Gan2011CITP,Xu2012FOP,Wang2022AMAC,shirsat2022PRE,Fei2023JOFM,
liang2024thermal,wang2017simulation,
shan2024morphological,zhuang2024deep,tao2017numerical,chen2024analysis,wang2020simple,wei2022small,fei2024pore,Hou2025POF}.

Among these methods, the LBM is recognized as a specialized discretized form of the Boltzmann equation. In 2003 and 2004, Watari \emph{et al.} and Kataoka \emph{et al.} developed LBMs that incorporate first-order TNEs and solve the time and space derivatives of the discrete Boltzmann equation using the finite difference (FD) method\cite{watari2003PRE,kataoka2004PRE}. This FD LBM abandons the traditional ``propagation + collision'' lattice gas model by decoupling velocity discretization from spatial discretization, thereby significantly enhancing the model's flexibility and applicability.
However, numerical instability limits the LBM's application to systems with Mach numbers greater than one. To address this limitation, various advancements have been proposed to extend its applicability in high-speed compressible flow simulations. These developments can be broadly classified into two main areas: physical modeling enhancements and numerical scheme improvements. Notable approaches include the dual-distribution-function scheme \cite{Wang2010WS,He2013E,Qiu2020AP,Qiu2021APS,Hosseini2024E}, the entropy principle method \cite{Frapolli2016APS}, the particle-on-demand method\cite{Kallikounis2022APS}, the recursive regularization scheme\cite{Feng2019E,Feng2020APS,Farag2020AP,coreixas2017recursive}, the multi-relaxation-time approach\cite{chen2010multiple,Safdari2019TF}, and advanced finite difference methods, such as the improved Lax-Wendroff FD scheme with additional viscosity\cite{Gan2008PA}, the fifth-order weighted essentially non-oscillatory FD scheme\cite{Gan2011APS}, the non-oscillatory, parameter-free dissipative scheme \cite{Chen2010IP}, the flux-limiting method\cite{Li2009E}, the implicit time-stepping approach\cite{Huang2024AP}, the flux solver scheme\cite{yang2020three}, and the finite volume method\cite{Yang2012CUP,Huang2023IP}, among others.
These improvements have significantly enhanced the stability and applicability of the LBM in high Mach number compressible flow simulations, offering a broader range of tools for studying complex hydrodynamic problems.

Despite these advancements, most studies primarily focus on first-order deviations of the distribution function from equilibrium, which limits their ability to fully capture and characterize transport behavior under strong nonequilibrium conditions. High-speed compressible flows further complicate this issue by exhibiting diverse flow patterns and complex mechanisms. In addition to the challenge of performing multiscale modeling, another critical issue lies in how to efficiently and intuitively extract nonequilibrium states, effects, and behaviors from the vast amounts of data generated by numerical simulations.

To address these issues, we resort to the discrete Boltzmann modeling and complex physical field analysis method (DBM)\cite{Xu2012FOP,Gan2015SM,Zhang2023CF,Xu2024FOP,Song2024POF,Chen2024SCPMA,Lai2024CF}. In 2012, Xu \textit{et al.} proposed a method that uses higher-order nonconserved kinetic moments of the distribution function to describe nonequilibrium states and extract associated TNE information\cite{Xu2012FOP}. Building on this foundation, Gan \textit{et al.} (2018) developed a multiscale DBM framework for high-speed compressible flows and derived second-order analytical expressions for key nonequilibrium quantities in the two-dimensional (2D) case\cite{Gan2018PRE}. In 2024, Xu \textit{et al.} further explored the ideas and strategies behind the DBM in coarse-grained modeling and complex physical field analysis, clarifying its relationship with traditional fluid modeling methods and other kinetic methods, and providing profound insights for its further development\cite{Xu2024FOP}.

Historically, the DBM evolved from the physical modeling branch of the LBM through selective taking, abandoning and supplementation. They start from the same point (kinetic theory), but aims at different requirements of the simulation study\cite{Xu2024FOP}.
As a discretized form of the simplified Boltzmann equation, DBM retains the use of discrete velocities but avoids relying on specific discrete formats, imposing only the essential physical constraints during discretization. These constraints include conserved kinetic moments and higher-order nonconserved kinetic moments, which are invariants in coarse-grained physical modeling and must be preserved. Such constraints provide a unique perspective for analyzing TNE behaviors, effects, and their manifestations.
As the order and variety of higher-order kinetic moment constraints increase, DBM goes beyond the continuity assumption and near-equilibrium approximations of traditional fluid modeling. Although the model's complexity increases slightly, its ability to describe nonequilibrium phenomena is significantly enhanced. Over time, DBM has developed new schemes for detecting, presenting, describing, and analyzing nonequilibrium states and effects based on the phase space description method and complex physical field analysis techniques\cite{Xu2024FOP}.
Specifically, each nonconserved kinetic moment provides a perspective on TNE intensity. These perspectives are interrelated and complementary, allowing for a more refined characterization of the physical structure and evolution of mesoscopic interfaces.

In recent years, DBM has been applied to a wide range of complex fluid systems, including multiphase flows\cite{Gan2015SM,Gan2022JOFM}, fluid instabilities\cite{Lai2016PRE,li2024kinetic}, detonation waves\cite{Lin2019PRE}, reactive flows\cite{zhang2016kinetic}, shock waves\cite{Qiu2020AP,Qiu2021APS}, micro–nanoscale flows\cite{zhang2023lagrangian}, and plasma kinetics\cite{Song2024POF}. These applications highlight the versatility of DBM in addressing multiscale and nonequilibrium problems.
For high-speed compressible flows in three dimensions, Gan \textit{et al.} proposed an efficient DBM with 30 discrete velocities to account for first-order TNEs\cite{Gan2018SPS}. Luo \textit{et al.} introduced a three-dimensional (3D) MRT DBM to capture HNE and TNE effects in reactive flows\cite{Ji2021AA}. In the same year, Luo \textit{et al.} also developed a 3D DBM to simulate both steady and unsteady detonation waves. This model improved the discrete velocity set by utilizing characteristic points of Platonic solids, achieving excellent spatial symmetry\cite{Ji2022JOCP}.

While these studies advanced the exploration of TNEs in 3D cases, they are limited to situations with weak nonequilibrium effects. In scenarios involving large gradients or intense TNEs, these models often fail to capture the finer details of the nonequilibrium phenomena. To address this limitation, extending DBM to the 3D Burnett level is essential for achieving a more comprehensive characterization of nonequilibrium properties.
In the field of higher-order DBM, Gan \textit{et al.} derived expressions for nonlinear viscous stress and heat flux by considering second-order TNEs using Chapman-Enskog (CE) analysis. They constructed a Burnett-level DBM to investigate HNEs and TNEs in high-speed compressible flows\cite{Gan2018PRE}. Later, they developed a multiscale, multiphase DBM based on density functional kinetic theory. This model, corresponding to the super-Burnett level, is suitable for studying TNEs in systems ranging from continuum to transition flow regimes\cite{Gan2022JOFM}.

Building upon previous work, this study constructs a 3D Burnett-level DBM to investigate HNE and TNEs in high-speed compressible flows. The core concept involves applying CE multiscale analysis to identify the kinetic moment relationships that should be preserved in coarse-grained physical modeling when considering second-order TNEs. Additionally, this study provides a theoretical derivation of nonlinear constitutive relations for the 3D case, analyzing how spatial dimensions influence the types of nonequilibrium driving forces and their coupling effects.
To facilitate this, a suitable discrete velocity set is designed to discretize the phase space, enabling the development of a robust and efficient 3D DBM.
The structure of the paper is organized as follows: Section \ref{Model} outlines the construction of the multiscale model. Section \ref{Numerical simulations} validates the model's accuracy in capturing large-scale structures in high-speed compressible flows, using standard test cases from 1D to 3D scenarios. Section \ref{TNE} evaluates the model's ability to describe higher-order TNEs across scales. Finally, Section \ref{Conclusions} concludes with a summary of the key findings.

\section{Three-dimensional high-order discrete Boltzmann model}\label{Model}

The complexity of 3D high-speed compressible flows arises from the diverse and persistent nonequilibrium driving forces that interact through competition and coordination, generating a broad spectrum of flow scales and kinetic modes. The DBM serves as a multiscale modeling and analysis tool with the following features:

(1) Modeling Stage: Using CE multiscale analysis, DBM efficiently identifies the kinetic moments required to describe TNEs. These kinetic moments are rigorously preserved during phase-space discretization, ensuring the model's ability to describe cross-scale phenomena.

(2) Analysis stage: By utilizing higher-order nonconserved kinetic moments $(f-f{^{(0)}})$, DBM enables the efficient extraction of nonequilibrium states, effects, and behaviors.  Moreover, constructing the phase space through the independent component of $\mathbf{M}_{m,n}(f-f{^{(0)}})$ offers a more intuitive representation of nonequilibrium states, effects, and behaviors.

The DBM modeling process involves linearizing the collision operator, identifying the kinetic moment relations needed to describe TNEs, discretizing particle velocity space, and extracting and characterizing nonequilibrium states, features, and effects. Details of these processes are described below.

\subsection{Simplification of the Boltzmann equation}

The Boltzmann equation is given as:
\begin{equation}\label{e1}
	\partial_t f+\mathbf{v} \cdot \bm{\nabla} f=J(f, f^*)=\iint(f^* f_1^*-f f_1) g \sigma d \Omega d \mathbf{v}_1,
\end{equation}
where \( f \) represents the particle velocity distribution function, and \( J(f, f^*) \) denotes the collision term, which involves the integral of collision contributions over particle velocities and solid angles.
The original collision term \( J(f, f^*) \) is highly complex and computationally challenging to solve.
To address this, the BGK model is introduced to simplify the collision term,
\begin{equation}\label{e2}
	\frac{\partial f}{\partial t}+\mathbf{v} \cdot \frac{\partial f}{\partial \mathbf{r}}=-\frac{1}{\tau}[f-f^{(0)}],
\end{equation}
where \( \tau \) is the relaxation time, and \( f^{(0)} \) represents the equilibrium distribution function.
In three dimensions, \( f^{(0)} \) is given by:
\begin{equation}\label{e3}
	f^{(0)} = \rho (\frac{1}{2 \pi R T})^{\frac{3}{2}} (\frac{1}{2 \pi n R T})^{\frac{1}{2}}
	\exp\left[-\frac{(\mathbf{v} - \mathbf{u})^2}{2 R T} - \frac{\eta^2}{2 n R T}\right],
\end{equation}
where \( \rho \), \( \mathbf{v} \), \( \mathbf{u} \), and \( T \) represent the local density, particle velocity, flow velocity, and temperature, respectively. The parameter \( R \) is the gas constant, and \( \eta \) is a free parameter accounting for the contribution of additional/internal degrees of freedom \( n \) (excluding translational motion).

\subsection{Determination of the essential kinetic moment relations and discretization of the particle velocity space}

To perform numerical simulation, Eq. (\ref{e2}) needs to be transformed into its discrete form, known as the discrete Boltzmann equation,
\begin{equation}\label{e4}
\frac{\partial f_i}{\partial t} + \mathbf{v}_i \cdot \frac{\partial f_i}{\partial \mathbf{r}}=-\frac{1}{\tau}[f_i-f_i^{(0)}],
\end{equation}
where \( i  = 1, \ldots, N \) represents the discrete velocity directions, \( f_i \) is the discrete distribution function, and \( f_i^{(0)} \) is the discrete equilibrium distribution function.
Discretizing the phase space involves dividing the continuous velocity space into a finite set of discrete points. Phase space discretization inevitably leads to ``loss of information''.
A key challenge in coarse-grained physical modeling is ensuring the preservation of essential information relevant to the numerical study during discretization.

In kinetic theory, the properties of a system are described by the distribution function and its kinetic moments. After discretization, the distribution function itself loses physical significance, while the integral or summation forms of the corresponding kinetic moments retain their physical relevance. Therefore, whether simplifying the collision term or discretizing the particle velocity space, the DBM must ensure that the fundamental physical constraint describing the system's behavior, represented by the kinetic moments, remains invariant, both before discretization in the integral form and after discretization in the summation form, as expressed below:
\begin{equation}\label{e5}
    \bm{\Phi}' = \int f \bm{\Psi}'(\mathbf{v}, \eta) d \mathbf{v} = \sum_i f_i \bm{\Psi}'(\mathbf{v}_i, \eta_i),
\end{equation}
where $\bm{\Psi}'(\mathbf{v}, \eta) = [ 1, \mathbf{v}, \frac{1}{2}(v^2 + \eta^2), \cdots, \mathbf{v} \mathbf{v} \mathbf{v} \mathbf{v}, \frac{1}{2}(v^2 + \eta^2) \mathbf{v} \mathbf{v} \mathbf{v}, . \\
	\quad . \mathbf{v} \mathbf{v} \mathbf{v} \mathbf{v} \mathbf{v}, \cdots ]$.

The CE multiscale analysis demonstrates that, recovering the hydrodynamic equations at different levels, requires satisfying specific kinetic moments. The DBM operates independently of the CE expansion. In DBM modeling, the CE multiscale analysis primarily aids in identifying and validating the kinetic moment relations retained in the coarse-grained model. The explicit derivation of macroscopic hydrodynamic equations is not a prerequisite for DBM modeling and simulations. To determine the required kinetic moments, the CE expansion is applied to both sides of Eq. (\ref{e4}), beginning with a multiscale expansion of the distribution function, time derivatives, and spatial derivatives:
\begin{equation}\label{e6}
    f_i = f_i^{(0)} + f_i^{(1)} + f_i^{(2)} + \cdots,
\end{equation}
\begin{equation}\label{e7}
    \partial_t = \partial_{t_1} + \partial_{t_2} + \cdots,
\end{equation}
\begin{equation}\label{e8}
    \bm{\nabla} = \bm{\nabla_1},
\end{equation}
where \( f_i^{(j)} \) represents the \( j \)th order deviation from \( f_i^{(0)} \), and \( \partial_{t_j} \) denotes the \( j \)th order term in the time scale expansion.

By substituting the above three expressions into Eq. (\ref{e4}) and combining terms of the same order, the expressions for $f_i^{(1)}$ and $f_i^{(2)}$ are obtained:
\begin{equation}\label{e9}
	f_i^{(1)} = -\tau \left[\partial_{t_1} f_i^{(0)} + \bm{\nabla}_1 \cdot (f_i^{(0)} \mathbf{v}_i)\right],
\end{equation}
\begin{equation}\label{e10}
	\begin{aligned}
		f_i^{(2)} = - & \tau \left[\partial_{t_2} f_i^{(0)} + \partial_{t_1}f_i^{(1)} + \bm{\nabla}_1 \cdot (f_i^{(1)} \mathbf{v}_i)\right] \\
		= - &\tau \partial_{t_2} f_i^{(0)} + \tau^2 \partial_{t_1}^2 f_i^{(0)} + \tau^2 \partial_{t_1} \left[\bm{\nabla}_1 \cdot (f_i^{(0)} \mathbf{v}_i)\right] \\
		+ & \tau^2 \bm{\nabla}_1 \cdot \left[\partial_{t_1} f_i^{(0)} \mathbf{v}_i + \bm{\nabla}_1 \cdot (f_i^{(0)} \mathbf{v}_i \mathbf{v}_i)\right].
	\end{aligned}
\end{equation}
It is evident that both \( f_i^{(1)} \) and \( f_i^{(2)} \) can be expressed in terms of \( f_i^{(0)} \). Specifically, \( f_i^{(1)} \) is represented by a first-order polynomial in \( \mathbf{v}_i \), while \( f_i^{(2)} \) is represented by a second-order polynomial in \( \mathbf{v}_i \). Consequently, to satisfy the kinetic moment relations for \( f_i^{(0)} \), the required accuracy must increase by one order for \( f_i^{(1)} \) and by two orders for \( f_i^{(2)} \).
Thus, Eq. (\ref{e5}) can be rewritten as:
\begin{equation}\label{e11}
    \bm{\Phi} = \int f^{(0)} \bm{\Psi}(\mathbf{v}, \eta) \, d \mathbf{v} = \sum_i f_i^{(0)} \bm{\Psi}(\mathbf{v}_i, \eta_i),
\end{equation}
where the term $\bm{\Psi}(\mathbf{v}, \eta)$ should include higher-order terms with powers greater than $\bm{\Psi}'(\mathbf{v}, \eta)$. Specifically, to account for the contributions of the second-order TNEs, $\bm{\Psi}(\mathbf{v}, \eta)=[1, \mathbf{v}, \frac{1}{2}(v^2+\eta^2), \mathbf{v v}, \frac{1}{2}(v^2+\eta^2)\mathbf{v}, \mathbf{v v v}, \frac{1}{2}(v^2+\eta^2) \mathbf{v v}, \mathbf{v v v v}, \frac{1}{2}(v^2+\eta^2)\mathbf{v v v}]$.
It is evident that the order and number of the preserved kinetic moments directly determine the modeling accuracy of the DBM, its ability to describe TNEs, and its cross-scale description capability.

The nine kinetic moments corresponding to Eq. (\ref{e11}) are as follows:
\begin{equation}\label{e12}
	\mathbf{M}_0=\sum_i f_i^{(0)}=\rho,
\end{equation}
\begin{equation}\label{e13}
	\mathbf{M}_1=\sum_i f_i^{(0)} \mathbf{v}_i=\rho \mathbf{u},
\end{equation}
\begin{equation}\label{e14}
	\mathbf{M}_{2,0}=\sum_i \frac{1}{2} f_i^{(0)}(v_i^2+\eta_i^2)=\frac{1}{2} \rho[(n+3) R T+u^2],
\end{equation}
\begin{equation}\label{e15}
	\mathbf{M}_2=\sum_i f_i^{(0)} \mathbf{v}_i \mathbf{v}_i=\rho(R T \mathbf{I}+\mathbf{u u}),
\end{equation}
\begin{equation}\label{e16}
	\mathbf{M}_{3,1}=\sum_i \frac{1}{2} f_i^{(0)}(v_i^2+\eta_i^2) \mathbf{v}_i=\frac{1}{2} \rho \mathbf{u}[(n+5) R T+u^2],
\end{equation}
\begin{equation}\label{e17}
	\begin{aligned}
		& \mathbf{M}_3=\sum_i f_i^{(0)} \mathbf{v}_i \mathbf{v}_i \mathbf{v}_i=\rho[R T (u_\alpha \delta_{\beta \gamma} \\
		&+ u_\beta  \delta_{\alpha \gamma}+  u_\gamma \delta_{\alpha \beta})\mathbf{e}_\alpha\mathbf{e}_\beta \mathbf{e}_\gamma+\mathbf{u u u}],
	\end{aligned}
\end{equation}
\begin{equation}\label{e18}
	\begin{aligned}
		\mathbf{M}_{4,2}= & \sum_i \frac{1}{2} f_i^{(0)}(v_i^2+\eta_i^2) \mathbf{v}_i \mathbf{v}_i=\frac{1}{2} \rho[(n+5) R^2 T^2 \\
		&+R T u^2] \mathbf{I}+\frac{1}{2} \rho[(n+7) R T+u^2] \mathbf{u u},
	\end{aligned}
\end{equation}
\begin{equation}\label{e19}
	\begin{gathered}
		\mathbf{M}_4=\sum_i f_i^{(0)} \mathbf{v}_i \mathbf{v}_i \mathbf{v}_i \mathbf{v}_i=\rho[R ^ { 2 } T ^ { 2 } (\delta_{\alpha \beta} \delta_{\gamma \lambda}+\delta_{\alpha \gamma} \delta_{\beta \lambda} \\
		+\delta_{\alpha \lambda} \delta_{\beta \gamma}) \mathbf{e}_\alpha \mathbf{e}_\beta \mathbf{e}_\gamma \mathbf{e}_\lambda+R T(u_\alpha u_\beta \delta_{\gamma \lambda}+u_\alpha u_\gamma \delta_{\beta \lambda}+u_\alpha u_\lambda \delta_{\beta \gamma} \\
		+u_\beta u_\gamma \delta_{\alpha \lambda}+u_\beta u_\lambda \delta_{\alpha \gamma}+u_\gamma u_\lambda \delta_{\alpha \beta}) \mathbf{e}_\alpha \mathbf{e}_\beta \mathbf{e}_\gamma \mathbf{e}_\lambda+\mathbf{u u u u}],
	\end{gathered}
\end{equation}
\begin{equation}\label{e20}
	\begin{aligned}
		& \mathbf{M}_{5,3}=\sum_i \frac{1}{2} f_i^{(0)}(v_i^2+\eta_i^2) \mathbf{v}_i \mathbf{v}_i \mathbf{v}_i = \rho \left[\left(\frac{n+9}{2} R T + \frac{u^2}{2}\right) \mathbf{u u u} \right. \\
		& \quad + \left.\left(\frac{n+7}{2} R T + \frac{u^2}{2}\right) R T (u_\alpha \delta_{\beta \gamma} + u_\beta \delta_{\alpha \gamma} + u_\gamma \delta_{\alpha \beta}) \mathbf{e}_\alpha \mathbf{e}_\beta \mathbf{e}_\gamma \right].
	\end{aligned}
\end{equation}

Equations (\ref{e12})-(\ref{e20}) can be expressed in matrix form as follows:
\begin{equation}\label{e21}
	\bm{\Phi} = \mathbf{C} \cdot \mathbf{f}^{(0)},
\end{equation}
$\bm{\Phi} = (\mathbf{M}_0, \mathbf{M}_1, \mathbf{M}_{2,0}, \mathbf{M}_{2},\mathbf{M}_{3,1},\mathbf{M}_{3}, \mathbf{M}_{4,2},\mathbf{M}_{4}, \mathbf{M}_{5,3})^T = (M_0, M_{1x}, M_{1y}, M_{1z}, \cdots, M_{5,3yzz},M_{5,3zzz})^T$
is a $55 \times 1$ column matrix. Each element of $\bm{\Phi}$ represents an independent kinetic moment of $f_i^{(0)}$.
The matrix $\mathbf{C} = (\mathbf{c}_1, \mathbf{c}_2, \cdots, \mathbf{c}_{55})$ is a $55 \times 55$ square matrix that establishes the relationship between the discrete equilibrium distribution function and the kinetic moments. Here, $\mathbf{c}_i$ is the $i$th column of $\mathbf{C}$,  defined as:
$\mathbf{c}_i = [1, v_{ix}, v_{iy}, v_{iz}, \cdots, \frac{1}{2}(v_i^2 + \eta_i^2) v_{iz} v_{iz} v_{iz}]^T$,
and $\mathbf{f}^{(0)} = (f_1^{(0)}, f_2^{(0)}, f_3^{(0)}, \cdots, f_{55}^{(0)})^T$ represents the vector of discrete equilibrium distribution functions.
Thus, $f_i^{(0)}$ can be computed as follows \cite{Gan2013EL}:
\begin{equation}\label{e22}
	\mathbf{f}^{(0)} = \mathbf{C}^{-1} \cdot \mathbf{M},
\end{equation}
where $\mathbf{C}^{-1}$ is the inverse of $\mathbf{C}$.

\begin{figure*}[htbp]
	\centering
	\includegraphics[width=0.92\textwidth]{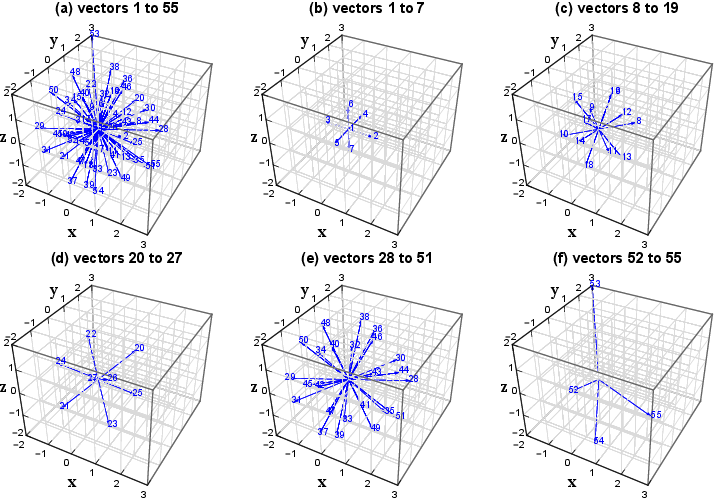}
	\caption{Schematic representation decomposition diagram of the D3V55 discrete velocity set.}
	\label{Fig01}
\end{figure*}

To efficiently handle the continuity of velocity space in numerical simulations, the originally continuous velocity space is discretized into a finite set of discrete velocities.
The design principles and optimization guidelines of the discrete velocity sets (DVSs) can be elaborated from the following four aspects\cite{Gan2018SPS,Gan2022JOFM,He2025POF}: (1) Determination of the number of discrete velocities: The minimum required number of discrete velocities is determined by the independent kinetic moment constraints that the distribution function must satisfy. In this study, these nine kinetic moments lead to 55 component equations, requiring at least 55 discrete velocities for accurate modeling. Increasing the number of discrete velocities improves the accuracy and stability of the model but reduces computational efficiency.

(2) Symmetry and asymmetry optimization:
The discrete velocity model is constructed based on symmetry principles, including geometric and physical symmetry.  Geometric symmetry ensures that discrete velocities are symmetrically distributed in space. Physical symmetry refers to the invariance of kinetic moment constraints under spatial rotation, requiring the moment matrix $\mathbf{C}$, constructed from discrete velocities, to be invertible. Consequently, a few asymmetric discrete velocities (e.g., velocities 52-55 in this study) are introduced.

(3) Discretization optimization strategies:
The effectiveness of DVSs can be evaluated by comparing numerical and analytical solutions for nonequilibrium effects. A well-designed DVS produces smooth and continuous $f_i$ while efficiently utilizing its more components for maintaining numerical robustness.\cite{He2025POF}

(4) Flexibility in discrete velocity model design:
DBM does not impose a fixed discrete velocity format. In practice, the DVS should be adjusted based on the problem characteristics and numerical feedback for optimal performance.

To ensure computational stability and accuracy, a 3D discrete velocity set with fifty-five velocities (D3V55), as shown in Fig. \ref{Fig01}(a), is proposed. The mathematical expression is as follows:
\begin{equation}\label{e23}
	(v_{i x}, v_{i y}, v_{i z}) = \begin{cases}
		(0, 0, 0) & \text{for } i = 1, \\
		cyc: c(\pm 1, 0, 0) & \text{for } 2 \leq i \leq 7, \\
		cyc: c(\pm 1, \pm 1, 0) & \text{for } 8 \leq i \leq 19, \\
		c(\pm 1, \pm 1, \pm 1) & \text{for } 20 \leq i \leq 27, \\
		cyc: c(\pm 1, \pm 2, 0) & \text{for } 28 \leq i \leq 51, \\
		c(3, -2, -1) \text{ or } c(-2, 1, 2) & \text{for } i = 52, \\
		c(-3, 2, 1) \text{ or } c(-2, 3, -2) & \text{for } i = 53, \\
		c(2, 1, -3) \text{ or } c(1, -2, -1) & \text{for } i = 54, \\
		c(-2, -1, 3) \text{ or } c(3, -2, 1) & \text{for } i = 55.
	\end{cases}
\end{equation}
Here, "$cyc$" represents cyclic permutation, which generates the corresponding vector by cyclically rearranging the digits and signs in the velocity vector. The decomposition diagram is shown in Fig. \ref{Fig01} (b)-(f). Specifically, (i) the first velocity corresponds to the zero velocity; (ii) the 2nd to 7th velocities correspond to the face-centered velocities; (iii) the eighth to nineteenth velocities correspond to the second diagonal velocities; (iv) the twentieth to twenty-seventh velocities correspond to the cube vertex velocities; (v) the twenty-eighth to fifty-first velocities correspond to the third diagonal velocities; and (vi) the final four velocities are special velocities with insufficient symmetry, used to ensure the invertibility of the velocity matrix.
We also set
\begin{equation}\label{e24}
\eta_i =
\begin{cases}
	 10 \eta_0 & \text{for } i = 1, \\
	 \eta_0 & \text{for } 2 \leq i \leq 7, \, i = 15, \, 19, \, 26,\\ & \hspace{1.3em} 35 \leq i \leq 40, \\
	 \eta_0 \text{ or } \eta_i = 0 & \text{for } 52 \leq i \leq 53, \\
	 0 & \text{for all other } i.
\end{cases}
\end{equation}
Here, \( c \) and \( \eta \) are two free parameters introduced to ensure the existence of the inverse matrix \( C^{-1} \) and to optimize the model’s stability and accuracy. The specific heat ratio is defined as \( \gamma = c_p/c_v = (n+5)/(n+3) \).

It should be noted that, the DBM, as a theoretical framework for model construction and complex physical field analysis, provides only the essential physical constraints required to describe the problem, namely equation \ref{e11}, without specifying temporal, spatial, or phase-space discretization schemes. The discrete velocity set presented here serves as an example to demonstrate the functionality of the DBM and may not necessarily be optimal.

\subsection{Derivation and extraction of thermodynamic nonequilibrium effects}

Taking the zeroth, first, and second-order moments of Eq. (\ref{e4}), yields the generalized hydrodynamic equations:
\begin{equation}\label{e25}
	\partial_t \rho + \bm{\nabla} \cdot(\rho \mathbf{u})=0,
\end{equation}		
\begin{equation}\label{e26}
	\partial_t \rho + \bm{\nabla} \cdot(\rho \mathbf{u u}+P \mathbf{I}+\bm{\Delta}_2^*)=0,
\end{equation}
\begin{equation}\label{e27}
	\partial_t(\rho e) + \bm{\nabla} \cdot[(\rho e+P) \mathbf{u}+\bm{\Delta}_2^* \cdot \mathbf{u}+\bm{\Delta}_{3,1}^*]=0,	
\end{equation}
where $P=\rho RT$ is the ideal gas equation of state, $e=c_v T+u^2/2$ denotes the energy density, and $c_v=(n+3) R/2$ is the specific heat at constant volume.

The CE multiscale analysis provides the theoretical foundation for such kinetic models. DBM extends beyond the generalized hydrodynamic equations from CE analysis in capturing system properties. From the perspective of kinetic macroscopic modeling, the physical functions of DBM correspond to the extended hydrodynamic equations, which encompass not only the evolution of conserved kinetic moments, such as mass, momentum, and energy, but also the evolution of closely related nonconserved kinetic moments. For example, evolution equations of $\bm{\Delta}_m^*$ and $\bm{\Delta}_{m,n}^*$ are\cite{Gan2022JOFM,Gan2025arXiv}
\begin{equation}
	\begin{aligned}
		& \partial_t \bm{\Delta}_m^*+\partial_t \mathbf{M}_m^*\left(f_i^{(0)}\right)+\boldsymbol{\nabla} \cdot\left[\mathbf{M}_{m+1}^*\left(f_i^{(0)}\right)\right. \\
		& \left.+\mathbf{M}_m^*\left(f_i^{(0)}\right) \mathbf{u}+\boldsymbol{\Delta}_{m+1}^*+\boldsymbol{\Delta}_m^* \mathbf{u}\right]=-\frac{1}{\tau} \boldsymbol{\Delta}_m^*,
	\end{aligned}
\end{equation}
\begin{equation}
	\begin{aligned}
		& \partial_t \boldsymbol{\Delta}_{m, n}^*+\partial_t \mathbf{M}_{m, n}^*\left(f_i^{(0)}\right)+\boldsymbol{\nabla} \cdot\left[\mathbf{M}_{m+1, n+1}^*\left(f_i^{(0)}\right)\right. \\
		& \left.+\mathbf{M}_{m, n}^*\left(f_i^{(0)}\right) \mathbf{u}+\boldsymbol{\Delta} _{m+1, n+1}^*+\boldsymbol{\Delta}_{m, n}^* \mathbf{u}\right]=-\frac{1}{\tau} \boldsymbol{\Delta}_{m, n}^* .
	\end{aligned}
\end{equation}

The evolution of low-order conserved kinetic moments is fundamental to hydrodynamics, while high-order nonconserved kinetic moments become more important with increased discretization and stronger nonequilibrium effects.

As the degree of nonequilibrium increases, the system's complexity rises sharply, necessitating additional physical quantities to accurately describe its state and behavior.
Traditional macroscopic modeling uses physical quantities, such as the Knudsen number, spatial gradients, and temporal variations, to characterize HNE effects.
DBM retains these conventional approaches while incorporating nonconserved kinetic moments of $(f_i-f_i^{(0)})$, which are explicitly defined as:
\begin{equation}\label{e28}
	\begin{aligned}
		& \bm{\Delta}_{m,n}=\mathbf{M}_{m,n}(f_i-f_i^{(0)})=\sum_i\left(\frac{1}{2}\right)^{1-\delta_{m n}} \\
		& (f_i-f_i^{(0)}) \underbrace{\mathbf{v}_i \mathbf{v}_i \cdots \mathbf{v}_i}_n({v}_i^2+\eta_i^2)^{\frac{m-n}{2}},
	\end{aligned}
\end{equation}
\begin{equation}\label{e29}
	\begin{aligned}
		& \bm{\Delta}_{m,n}^*=\mathbf{M}_{m, n}^*(f_i-f_i^{(0)})=\sum_i\left(\frac{1}{2}\right)^{1-\delta_{m n}} \\
		& (f_i-f_i^{(0)}) \underbrace{\mathbf{v}_i^* \mathbf{v}_i^* \cdots \mathbf{v}_i^*}_n({v}_i^{* 2}+\eta_i^{2})^{\frac{m-n}{2}},
	\end{aligned}
\end{equation}
where $\delta_{m n}$ is the Kronecker delta function, and $\mathbf{v}_i^*=\mathbf{v}_i-\mathbf{u}$, $\bm{\Delta}_{m n}^*$ represents the contraction of an $m$th order tensor to an $n$th order one.
When $m=n$, $\bm{{\Delta}}_{m,n}^*$ is abbreviated as $\bm{{\Delta}}_{m}^*$.
If \( \mathbf{v}_i^* \) is replaced by \( \mathbf{v}_i \), the central moment \( \bm{\Delta}_{m,n}^* \) transforms into the non-central moment  \( \bm{\Delta}_{m,n} \).
The central moment \( \bm{\Delta}_{m,n}^* \) captures the thermal fluctuations of microscopic particles relative to $\mathbf{u}$, describing only TNEs. In contrast, \(\bm{\Delta_{m,n}} \) includes both flow velocity and thermal fluctuations, capturing both HNE and TNE effects.
These two descriptions are interconnected and complementary, providing unique perspectives for describing or measuring the system's deviation from thermodynamic equilibrium. Additionally, the independent components of
 \( \bm{\Delta_{m,n}} \) and \( \bm{\Delta_{m,n}}^* \)
can form the basis for constructing a phase space, offering a geometric framework for representing the states and behavior of complex systems.
The use of nonconserved kinetic moments of $(f_i-f_i^{(0)})$ to detect and describe deviation from thermodynamic equilibrium, along with their associated effects, represents a key feature of DBM in complex physical field analysis.

Two of the most critical nonequilibrium quantities in fluid modeling are the viscous stress ${\bm{\Delta}}_{2}^*$ and heat flux ${\bm{\Delta}}_{3,1}^*$, for which we derive analytical expressions using the CE multiscale expansion.

Taking the velocity moments with the collision-invariant vector $\left( 1, \mathbf{v}_i, (v_i^2 + \eta_i^2)/2 \right)$ on both sides of Eq.~(\ref{e9}), we obtain the following relations between the temporal derivative $\partial_{t_1}$ and the spatial derivative $\bm{\nabla}_1$:
\begin{equation}\label{eA1}
\partial_{t_1} \rho = -\bm{\nabla}_1 \cdot (\rho \mathbf{u}),
\end{equation}
\begin{equation}\label{eA2}
\partial_{t_1} \mathbf{u} = -R \bm{\nabla}_1 T - \frac{R T}{\rho} \bm{\nabla}_1 \rho - \mathbf{u} \cdot \bm{\nabla}_1 \mathbf{u},
\end{equation}
\begin{equation}\label{eA3}
\partial_{t_1} T = -\mathbf{u} \cdot \bm{\nabla}_1 T - \frac{2 T}{n+3} \bm{\nabla}_1 \cdot \mathbf{u}.
\end{equation}

Using Eqs.~(\ref{eA1})–(\ref{eA3}) and performing algebraic manipulation, we obtain the first-order viscous stress and heat flux:
\begin{equation}\label{D2}
   	\bm{\Delta}_2^{*(1)} = \sum_i f_i^{(1)} \mathbf{v}_i^* \mathbf{v}_i^*
    	= -\mu \left[ \bm{\nabla} \mathbf{u}
	+ \left( \bm{\nabla} \mathbf{u} \right)^T - \frac{2}{n+3} \mathbf{I} \bm{\nabla} \cdot \mathbf{u} \right]
	= -\bm{\sigma}_\text{NS},
\end{equation}
\begin{equation}\label{D31}
	\bm{\Delta}_{3,1}^{*(1)}=\sum_i \frac{1}{2} f_i^{(1)} \frac{v_i^{* 2}+\eta_i^{2}}{2} \mathbf{v}_i^*=-\kappa \bm{\nabla} T=-\mathbf{j}_{q, \text{NS}},
\end{equation}
where $\mu = P \tau$, $\kappa = \mathrm{c}_P P \tau$, and $\mathrm{c}_P = (n+5) R/2$ represent the viscosity coefficient, heat conductivity coefficient, and specific heat at constant pressure, respectively.

Similarly, by applying the collision-invariant vector to both sides of Eq.~(\ref{e10}), we obtain relations between the temporal derivative $\partial_{t_2}$ and the spatial derivative $\bm{\nabla}_1$:
\begin{equation}\label{eA4}
	\partial_{t_2} \rho=0,
\end{equation}
\begin{equation}\label{eA5}
	\rho \partial_{t_2} \mathbf{u}=-\bm{\nabla}_1 \cdot \bm{\Delta}_2^{(1)},
\end{equation}
\begin{equation}\label{eA6}
	\partial_{t_2}\big[\frac{(n+3)}{2} \rho T+\frac{1}{2} \rho u^2\big]=-\bm{\nabla}_1 \cdot \bm{\Delta}_{3,1}^{(1)}.
\end{equation}
Using Eqs.~(\ref{eA4})–(\ref{eA6}), we obtain the second-order viscous stress and heat flux, detailed in Appendix \ref{App}:
\begin{equation}\label{e40}
	\bm{\Delta}_2^{*(2)}=\sum_i  f_i^{(2)} \mathbf{v}_i^* \mathbf{v}_i^*=-(\bm{\sigma}_{\text {Burnett }}-\bm{\sigma}_{\text{NS}}), \\
\end{equation}
\begin{equation}\label{e41}
	\bm{\Delta}_{3,1}^{*(2)}=\sum_i  f_i^{(2)} \frac{v_i^{* 2}+\eta_i^{2}}{2} \mathbf{v}_i^*=-(\mathbf{j}_{q, \text {Burnett}}-\mathbf{j}_{q, \text{NS}}).
\end{equation}

These equations provide explicit expressions for the second-order constitutive relations. These equations can not only improve traditional macroscopic fluid modeling but also guide cross-scale regulation of nonequilibrium effects. Here, $(\bm{\sigma}_{\text{Burnett}} - \bm{\sigma}_{\text{NS}})$ represents the components where the NS equations fail to describe the system within the range of applicability of the Burnett equations. Therefore, the second-order constitutive relations are given by:
\begin{equation}\label{e34}
	\bm{\Delta}_2^*=\bm{\Delta}_2^{*(1)}+\bm{\Delta}_2^{*(2)},
\end{equation}
\begin{equation}\label{e35}
	\bm{\Delta}_{3,1}^*=\bm{\Delta}_{3,1}^{*(1)}+\bm{\Delta}_{3,1}^{*(2)}.
\end{equation}

Other higher-order thermodynamic nonequilibrium quantities can also be analytically derived  similarly, for example, $\Delta_{3 x x x}^{*(1)}=-3 \tau \rho R^2 T {\partial x} T$, $\Delta_{3 x x x}^{*(2)}=6 n_3^{-1} \tau^2 \rho R^2 T\Big[n_1 T \frac{\partial^2}{\partial x^2} u_x-2 T \frac{\partial^2}{\partial x \partial y} u_y-2 T \frac{\partial^2}{\partial x \partial z} u_z +\partial_x T\big((3 n+5) \partial_x u_x-4 \partial_y u_y-4 \partial_z u_z\big)+n_3\big(\partial_y T \partial_y u_x+\partial_z T \partial_z u_x\big)\Big]$.

Meanwhile, we point out that using a single relaxation time in the collision term fixes the Prandtl number at $\Pr = 1$.
To overcome this limitation, an external forcing term, $I_i = [A R T + B(\mathbf{v}_i - \mathbf{u})^2] f_i^{(0)}$, is added to the right-hand side of Eq.~(\ref{e4}) to modify the BGK collision operator \cite{Gan2011CITP}. Here, $A = -3 B$ and $B = \frac{1}{3 \rho T^2} \bm{\nabla} \cdot \left[ \frac{5+n}{2} \rho T q \bm{\nabla} T \right]$. Consequently, the heat conductivity becomes $\kappa = c_p P(\tau + q)$, and the Prandtl number is updated as $\Pr = \tau / (\tau + q)$. This method is straightforward and eliminates the need for $f^{(0)}$ to satisfy additional kinetic moments.
We also point out that, this approach is based on the equilibrium distribution function. It allows for the introduction of HNEs, but does not introduce additional TNEs.
To further improve the model, advanced schemes such as the Shakhov model, elliptic statistical model, multi-relaxation time, or two-relaxation time approaches can be considered.

\section{ Numerical tests and analysis }\label{Numerical simulations}

In this section, several typical test cases, ranging from 1D to 3D, are used to evaluate the model's capability to capture large-scale flow structures. To ensure high accuracy, stability, and computational efficiency, we employ the following numerical schemes: the time derivative is discretized using a second-order implicit-explicit Runge-Kutta scheme \cite{ascher1997ANM}, and the spatial derivative is approximated using either a fifth-order weighted essentially non-oscillatory (fifth WENO) scheme \cite{liu1994JOCP} or a second-order non-oscillatory, parameter-free dissipative (second NND) scheme \cite{Zhang1988}. As the extent of nonequilibrium effects increases, higher-order numerical schemes become necessary to maintain numerical solution accuracy and stability.
The discrete Boltzmann equation, particle velocity, and hydrodynamic quantities are nondimensionalized using appropriate reference variables \cite{Gan2011APS}. The three independent reference variables are the characteristic flow length scale $L_0$, the reference density $\rho_0$, and the reference temperature $T_0$.
Consequently, dimensionless variables are
$\hat{r}_{\alpha} = \frac{r_\alpha}{L_0}$, $( \hat{v}_\alpha, \hat{c}_\alpha, \hat{u}_\alpha ) = \frac{(v_\alpha, c_\alpha, u_\alpha)}{\sqrt{R T_0}}$, $\hat{\rho} = \frac{\rho}{\rho_0}$, $\hat{T} = \frac{T}{T_0}$,
$(\hat{\tau},\hat{t}) = \frac{(\tau,t)}{L_0 / \sqrt{R T_0}}$, and $\left(\hat{f}, \hat{f}^{(0)}, \hat{f}_i, \hat{f}_i^{(0)}\right)=\frac{\left(f, f^{(0)}, f_i, f_i^{(0)}\right)}{\rho_0\left(R T_0\right)^{-3/2}}$.
For simplicity, the symbol ``$\wedge$'' associated with dimensionless variables will be omitted henceforth.

\subsection{1D Riemann problems}

In this section, the one-dimensional Riemann problem is employed to validate the model’s ability to describe large-scale structures. Specifically, Riemann problems, such as Sod and Lax shock tubes, involve mesoscale structures—shock waves, rarefaction waves, and contact discontinuities—where significant nonequilibrium effects arise. These problems also provide an opportunity to assess the model’s capability in describing small-scale structures and its ability to capture the multiscale nature of thermodynamic nonequilibrium effects.

\subsubsection{Einfeldt shock tube}

\begin{figure}[htbp]
	{\centering
		\includegraphics[width=0.5\textwidth]{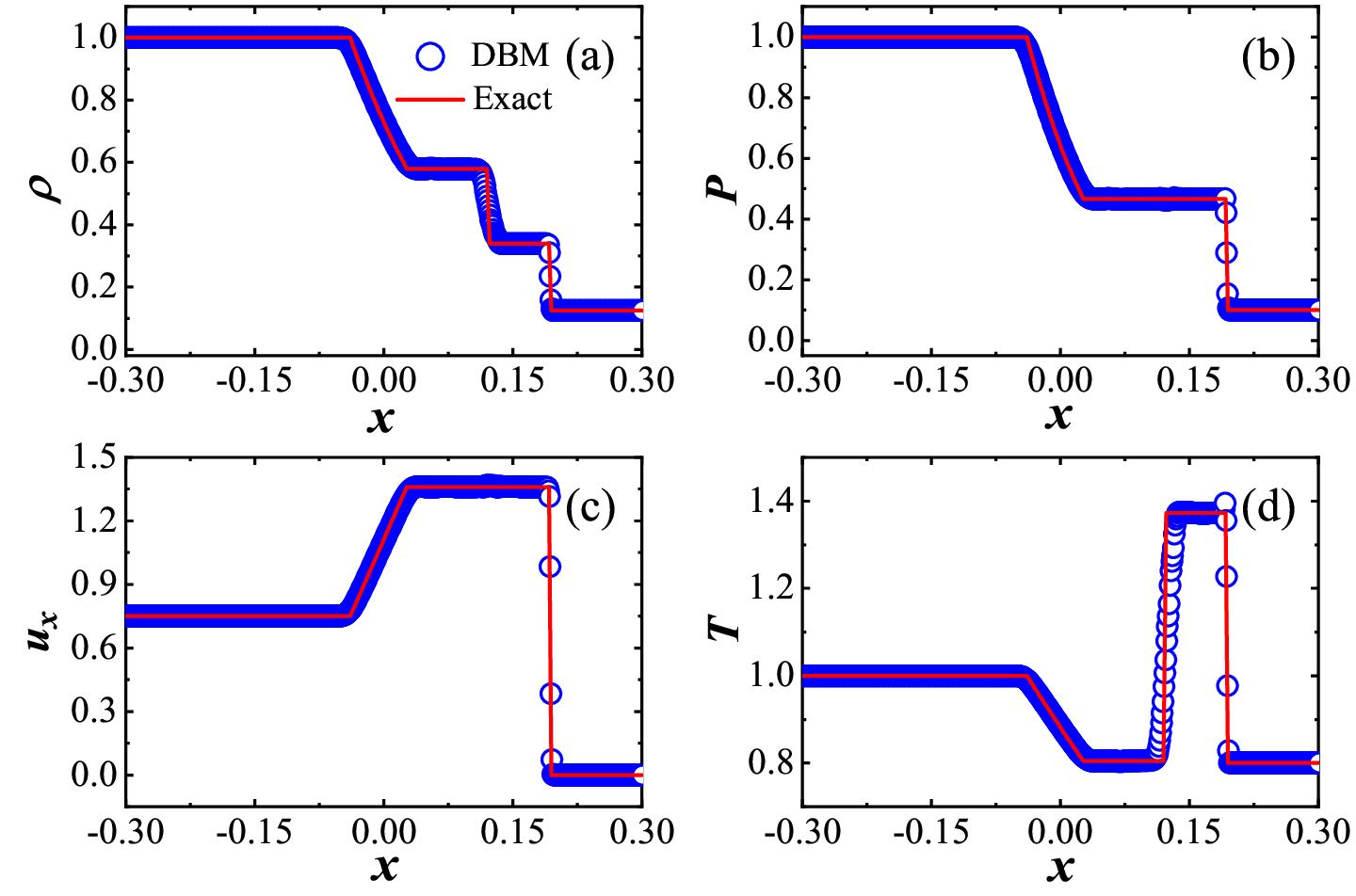}}
	\caption{\centering{Comparisons between DBM simulations and Riemann solutions for the Einfeldt shock tube at $t = 0.1$: (a) density, (b) pressure, (c) velocity, and (d) temperature.}}
	\label{Fig02}
\end{figure}

The Einfeldt shock tube problem, proposed by Einfeldt in 1988, is a classical benchmark widely used to evaluate the performance of numerical methods\cite{einfeldt1988SJONA}. Its analytical solution consists of three characteristic structures: a rightward propagating shock, a leftward propagating rarefaction wave, and a contact discontinuity between them. The initial conditions are given as:
\begin{equation}\label{e40}
    \left\{
    \begin{array}{l}
        \left( \rho, T, u_x, u_y, u_z \right)|_L = \left( 1.0, 1.0, 0.75, 0.0, 0.0 \right), \\
        \left( \rho, T, u_x, u_y, u_z \right)|_R = \left( 0.125, 0.8, 0.0, 0.0, 0.0 \right).
    \end{array}
    \right.
\end{equation}
Here, ``$L$'' and ``$R$'' denote the initial macroscopic quantities on the left and right sides of the discontinuity, respectively. Unlike the Sod shock tube problem, the Einfeldt shock is driven by not only the initial differences in pressure and density but also the initial flow velocity.
The computational grid is defined as \( N_x \times N_y \times N_z = 1000 \times 3 \times 3 \), with a spatial resolution of \( \Delta x = \Delta y = \Delta z = 10^{-3} \) and a time step size of \( \Delta t = 10^{-4} \). The model parameters are set as \( \tau = 10^{-4} \), \( n = 2 \), \( c = 1.5 \), and \( \eta_0 = 4.7 \). Periodic boundary conditions are applied in the \( y \)- and \( z \)-directions, while the boundary conditions in the \( x \)-direction are specified as follows. At the left boundary,
\begin{equation}
    f_{i,-1,t} = f_{i,0,t} = f_{i,1,t=0}^{(0)},
\end{equation}
indicating that the system remains in equilibrium at the boundary and the macroscopic quantities satisfy:
\begin{equation}
    (\rho, T, u_x, u_y, u_z)|_{-1,t} = (\rho, T, u_x, u_y, u_z)|_{0,t} = (\rho, T, u_x, u_y, u_z)|_{1,t=0}.
\end{equation}
At the right boundary,
\begin{equation}
    f_{i,N_x+2,t} = f_{i,N_x+1,t} = f_{i,N_x,t=0}^{(0)},
\end{equation}
ensuring equilibrium with the macroscopic quantities,
\begin{equation}
\begin{aligned}
(\rho, T, u_x, u_y, u_z)|_{N_x+2,t} &= (\rho, T, u_x, u_y, u_z)|_{N_x+1,t} \\
&= (\rho, T, u_x, u_y, u_z)|_{N_x,t=0}.
\end{aligned}
\end{equation}

Figure \ref{Fig02} compares the DBM numerical results with the Riemann analytical solution at \( t = 0.1 \). The blue hollow circles represent the numerical solution, while the red solid line corresponds to the analytical solution. The DBM accurately captures key features, such as the shock wave, expansion wave, and contact discontinuity, showing excellent agreement with the analytical solution. Additionally, numerical dissipation and oscillations are effectively suppressed, demonstrating the robustness and accuracy of the DBM and its numerical method.

\subsubsection{Modified Sod shock tube}

\begin{figure}[htbp]
	{\centering
		\includegraphics[width=0.5\textwidth]{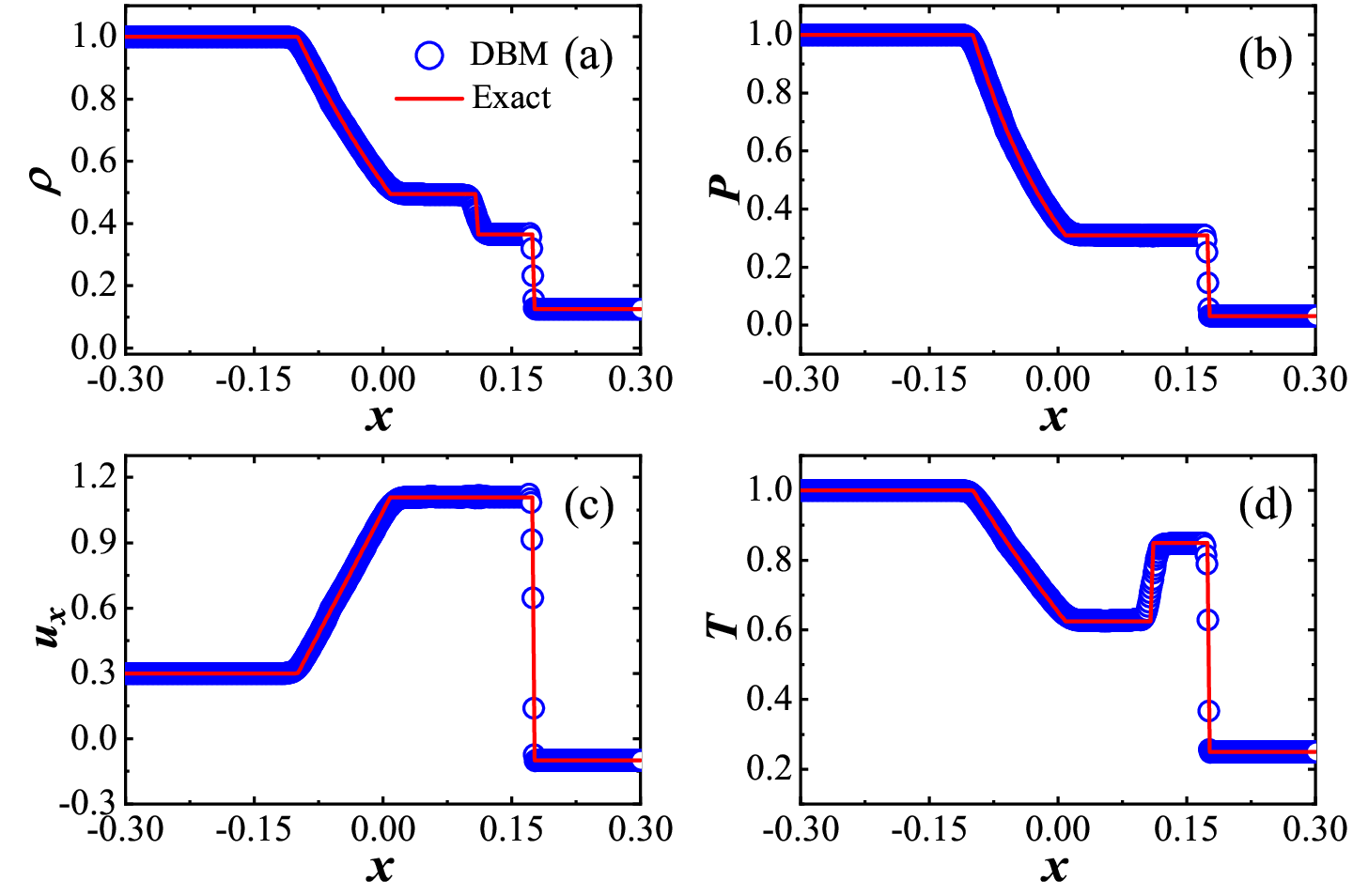}}
	\caption{\centering{Comparisons between DBM simulations and Riemann solutions for the modified Sod shock tube at $t = 0.1$: (a) density, (b) pressure, (c) velocity, and (d) temperature.}}
	\label{Fig03}
\end{figure}

The Sod shock tube problem, first introduced by Gary A. Sod in 1978, serves as a classical benchmark for evaluating compressible flow simulation models \cite{sod1978JOCP}. To further validate the model's robustness in more complex scenarios, a modified version of the Sod shock tube has been designed. This modification builds upon the original Sod shock tube by lowering the temperature on the right side of the discontinuity, thereby increasing the pressure ratio. Additionally, an initial velocity of $u_{x}=0.3$ is assigned to the fluid on the left side, while $u_{x}=-0.1$ is applied to the fluid on the right side. These changes enhance the velocity difference, providing a rigorous test of the model’s capability to handle larger temperature, pressure, and velocity gradients. The initial conditions are specified as follows:
\begin{equation}\label{e41}
    \left\{
    \begin{array}{l}
        \left( \rho, T, u_x, u_y, u_z \right)|_L = \left( 1.0,1.0,0.3,0.0,0.0 \right), \\
        \left( \rho, T, u_x, u_y, u_z \right)|_R = \left( 0.125,0.25,-0.1,0.0,0.0 \right).
    \end{array}
    \right.
\end{equation}
Here, $\tau=6 \times 10^{-5}, n=0, c=1.2, \eta_0=4.9$, while the remaining parameters remain unchanged.
Figure \ref{Fig03} presents a comparison between the DBM numerical solution and the Riemann analytical solution for the macroscopic quantities of the 1D modified Sod shock tube at $t = 0.1$. The results show excellent agreement between the two methods. The DBM demonstrates high accuracy in capturing complex wave structures, including shock waves, contact discontinuities, and rarefaction waves, with generally small errors. This highlights the model's adaptability, accuracy, and stability in simulating compressible flows with complex initial conditions.

\subsubsection{Lax shock tube}

\begin{figure}[htbp]
	{\centering
		\includegraphics[width=0.5\textwidth]{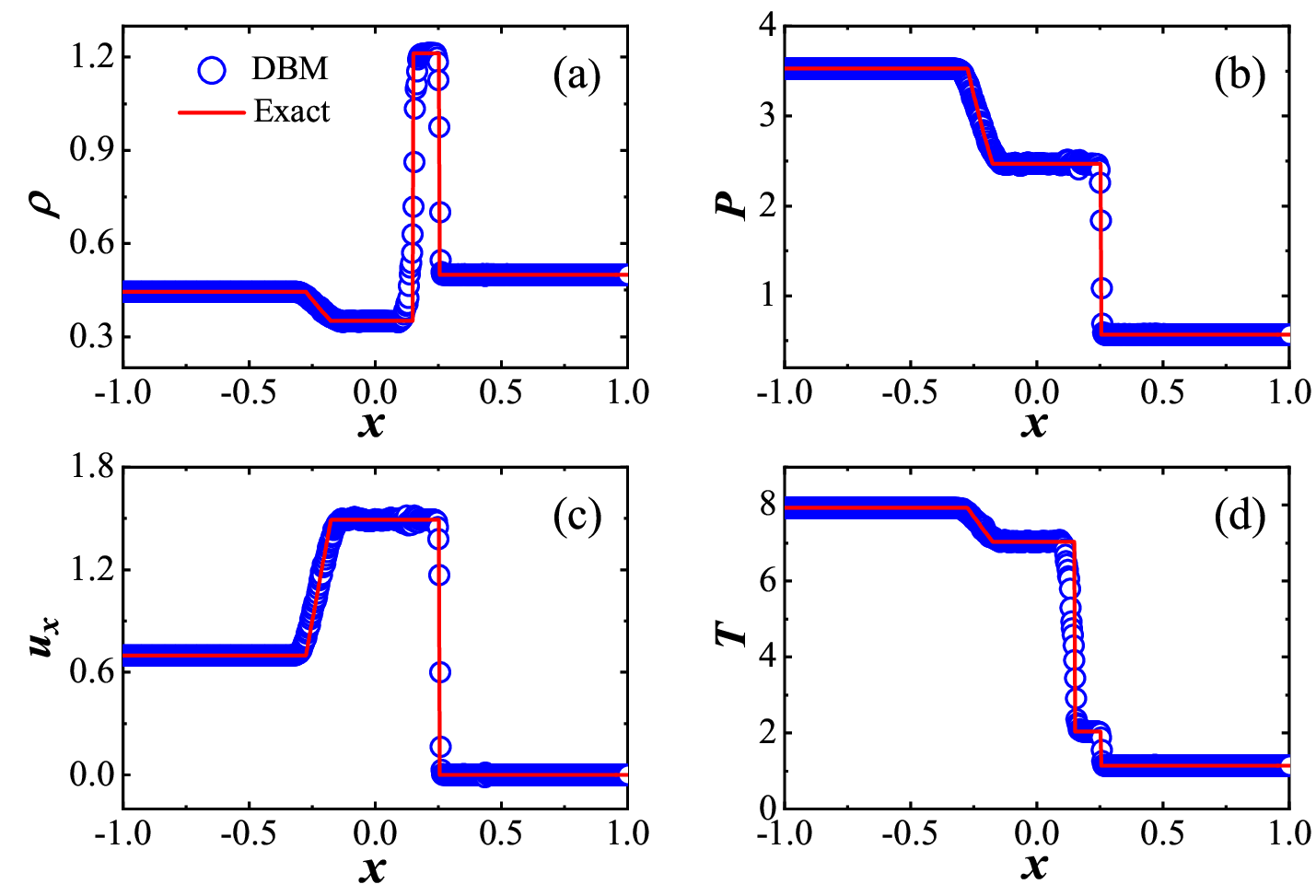}}
	\caption{\centering{Comparisons between DBM simulations and Riemann solutions for the Lax shock tube at $t = 0.1$: (a) density, (b) pressure, (c) velocity, and (d) temperature.}}
	\label{Fig04}
\end{figure}

The third test is the Lax shock tube problem, with the specific initial conditions given as follows:
\begin{equation}\label{e42}
    \left\{
    \begin{array}{l}
        \left( \rho, T, u_x, u_y, u_z \right)|_L = \left( 0.445,7.928,0.698,0.0,0.0 \right), \\
        \left( \rho, T, u_x, u_y, u_z \right)|_R = \left( 0.50,1.142,0.0,0.0,0.0\right).
    \end{array}
    \right.
\end{equation}
Here $\Delta x=\Delta y=\Delta z=3 \times 10^{-3}, \tau=3 \times 10^{-5}, n=1, c=-2.6, \eta_0=12.8$, while the other parameters remain unchanged.
Figure \ref{Fig04} compares the DBM numerical solution with the Riemann analytical solution, demonstrating excellent agreement and validating the model's effectiveness. The results owns the following key features. The density gradually decreases in the expansion wave region and exhibits a jump at the contact discontinuity. The pressure decreases in the expansion wave region and remains constant at the contact discontinuity. The velocity gradually increases in the expansion wave region and remains constant at the contact discontinuity. The temperature decreases in the expansion wave region, forming a temperature gradient at the contact discontinuity.
In the shock wave region, the density, pressure, velocity, and temperature all increase sharply. Numerical dissipation is significantly suppressed, highlighting the model's capability to handle compressible flows and complex wave structures with high stability and accuracy.

\subsubsection{Collision of two strong shocks}

\begin{figure}[htbp]
	{\centering
		\includegraphics[width=0.5\textwidth]{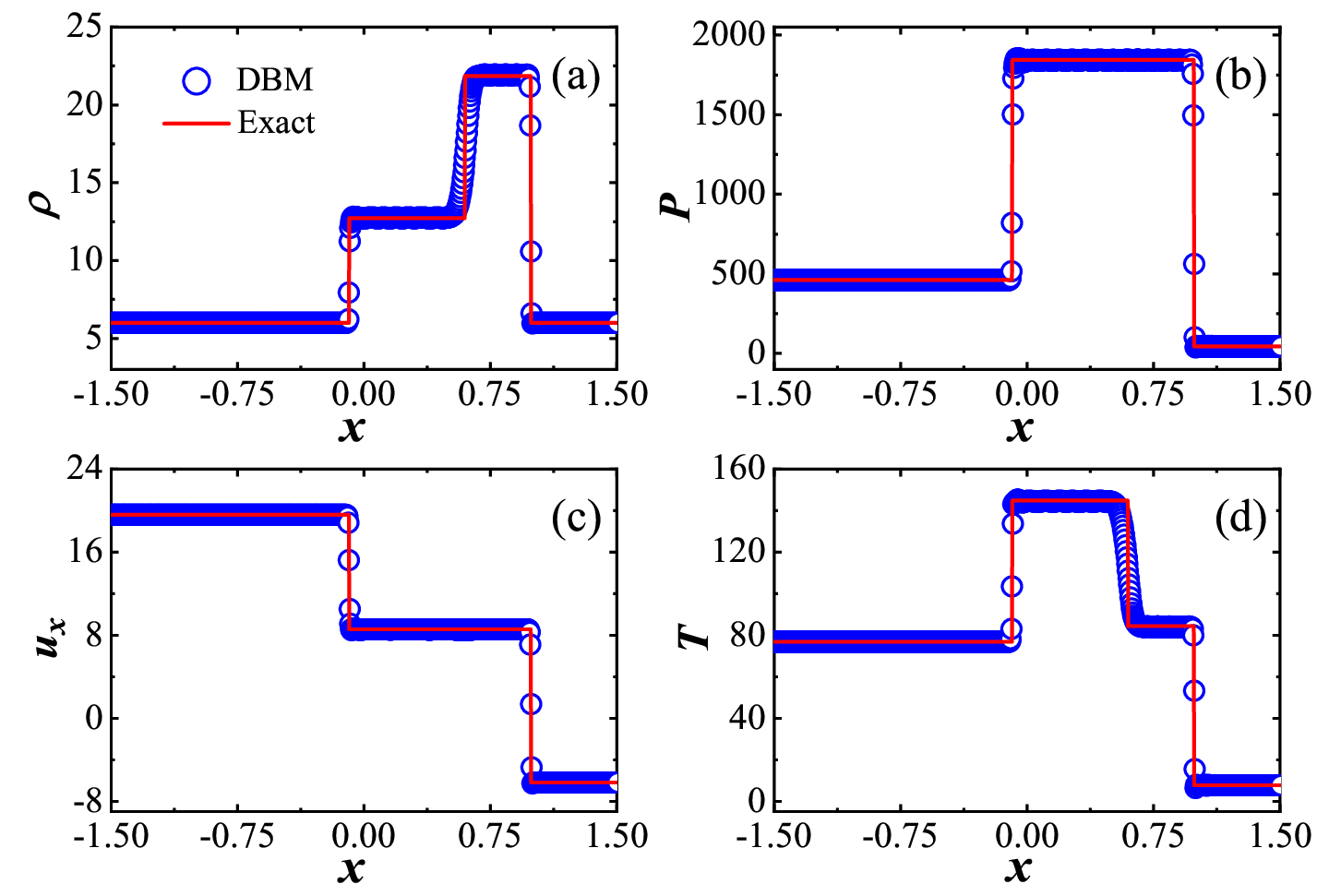}}
	\caption{\centering{Comparisons between DBM simulations and Riemann solutions for the collision of two strong shocks at $t = 0.1$: (a) density, (b) pressure, (c) velocity, and (d) temperature.}}
	\label{Fig05}
\end{figure}

Finally, a more challenging shock tube problem involving strong shock wave collisions is selected. This scenario also includes key features such as shock waves, expansion waves, and contact discontinuities. The specific initial conditions are given as follows:
\begin{equation}\label{e43}
    \left\{
    \begin{array}{l}
        \left( \rho, T, u_x, u_y, u_z \right)|_L = \left( 5.99924,76.8254,19.5975,0.0,0.0 \right), \\
        \left( \rho, T, u_x, u_y, u_z \right)|_R = \left( 5.99242,7.69222,-6.19633,0.0,0.0 \right).
    \end{array}
    \right.
\end{equation}
The numerical parameters for this problem are $\Delta x = \Delta y = \Delta z = 4 \times 10^{-3}$, $\Delta t = 5 \times 10^{-5}$, $\tau = 5 \times 10^{-5}$, $n = 0$, $c = -13$, and $\eta_0 = 34$, while the remaining parameters remain unchanged.
The numerical solutions for density, pressure, velocity, and temperature distributions closely align with the analytical solutions.
The model handles strong shock waves with remarkable stability and accuracy, exhibiting minimal numerical dissipation [see Fig. \ref{Fig05}].
It is worth noting that slight diffusion is observed in the steep shock wave regions. This diffusion is attributed to the DBM’s inherent dissipative mechanisms, such as viscous stress and heat conduction, enabling more effective representation of meso-scale structures like shock interfaces. This confirms the model's robustness and stability in handling strong nonlinear compressible flows.

\subsection{2D Riemann problems}

\begin{figure*}[htbp]
	{\centering
		\includegraphics[width=0.72\textwidth]{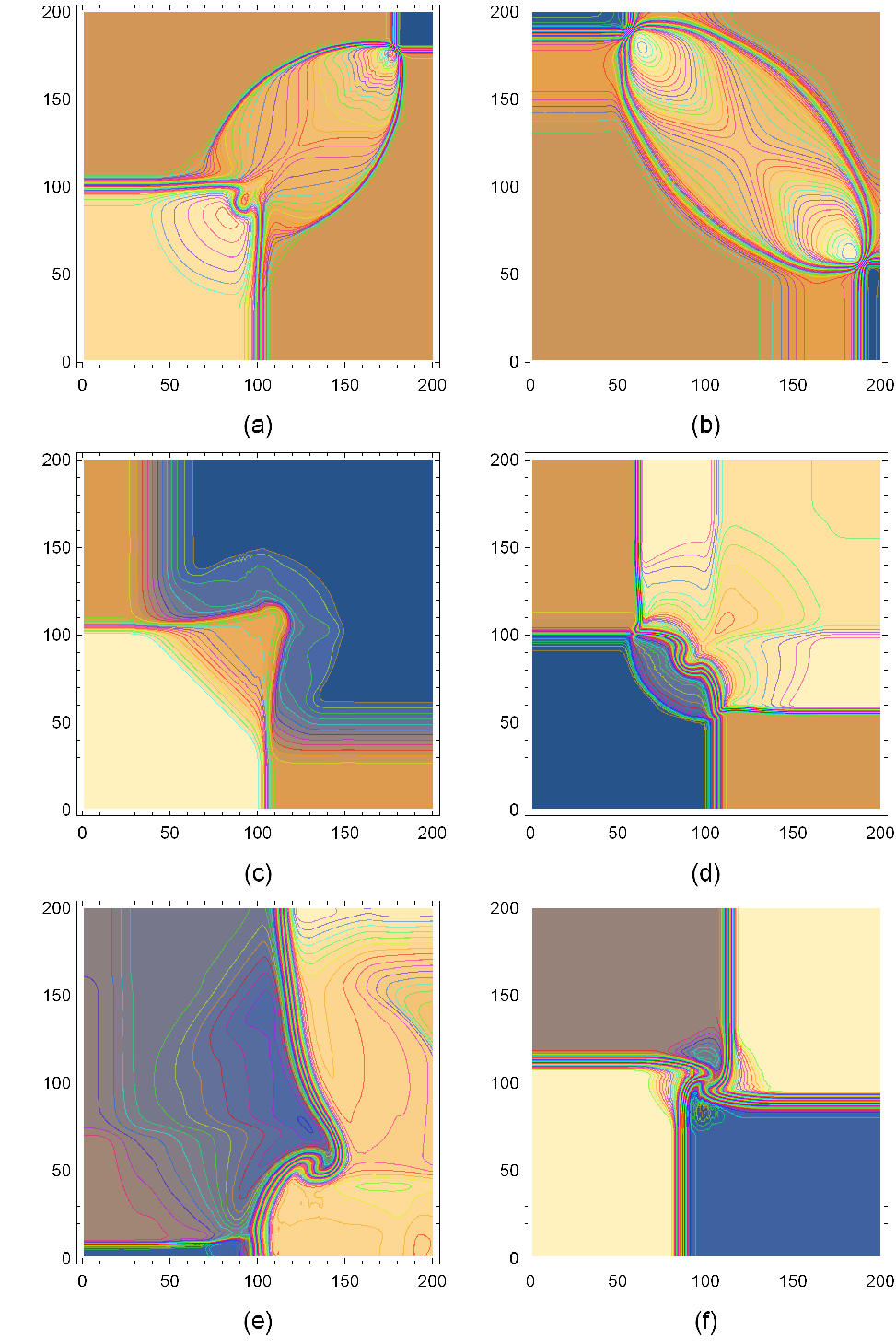}}
	\caption{\centering{Density distribution maps for the 2D Riemann problems calculated from the D3V55 model under six distinct configurations.}}
	\label{Fig06}
\end{figure*}

%%%%%%%%%%%%%%%%%%%%%%%%%%%%%%%%%%%%%%%%%%%%%%%%%%%%%%%%%%%%%%%%%%%%
\begin{table*}[htbp]
	\centering
	\caption{Initial conditions for the 2D Riemann problems.}
	\begin{tabular}{c c c}
		\toprule
		Configuration &
		\(\left(\begin{array}{c c|c c}
			P_2 & \rho_2 & P_1 & \rho_1 \\
			u_{x2} & u_{y2} & u_{x1} & u_{y1} \\
			\hline
			P_3 & \rho_3 & P_4 & \rho_4 \\
			u_{x3} & u_{y3} & u_{x4} & u_{y4} \\
		\end{array}\right)\) \\
		\midrule
		
		\multirow{2}{*}{Configurations (a) and (b)} &
		\(\left(\begin{array}{c c|c c}
			1 & 1 & 0.4 & 0.5313 \\
			0.7276 & 0 & 0 & 0 \\
			\hline
			1 & 0.8 & 1 & 1 \\
			0 & 0 & 0 & 0.7276 \\
		\end{array}\right)\) &
		\(\left(\begin{array}{c c|c c}
			0.35 & 0.5065 & 1.1 & 1.1 \\
			0.8939 & 0 & 0 & 0 \\
			\hline
			1.1 & 1.1 & 0.35 & 0.5065 \\
			0.8939 & 0.8939 & 0 & 0.8939 \\
		\end{array}\right)\) \\
		
		\midrule
		\multirow{2}{*}{Configurations (c) and (d)} &
		\(\left(\begin{array}{c c|c c}
			1 & 1 & 0.4 & 0.5197 \\
			-0.6259 & 0.1 & 0.1 & 0.1 \\
			\hline
			1 & 0.8 & 1 & 1 \\
			0.1 & 0.1 & 0.1 & -0.6259 \\
		\end{array}\right)\) &
		\(\left(\begin{array}{c c|c c}
			0.4 & 0.5313 & 1 & 1 \\
			0.8276 & 0 & 0.1 & 0 \\
			\hline
			0.4 & 0.8 & 0.4 & 0.5313 \\
			0.1 & 0 & 0.1 & 0.7276 \\
		\end{array}\right)\) \\
		
		\midrule
		\multirow{2}{*}{Configurations (e) and (f)} &
		\(\left(\begin{array}{c c|c c}
			1 & 2 & 1 & 1 \\
			0 & -0.3 & 0.3 & 0 \\
			\hline
			0.4 & 1.0625 & 0.4 & 0.5197 \\
			0 & 0.2145 & 0 & -0.4259 \\
		\end{array}\right)\) &
		\(\left(\begin{array}{c c|c c}
			1 & 2 & 1 & 1 \\
			0.5 & 0.5 & 0.5 & -0.5 \\
			\hline
			1 & 1 & 1 & 3 \\
			-0.5 & 0.5 & -0.5 & -0.5 \\
		\end{array}\right)\) \\
		
		\bottomrule
	\end{tabular}
	\label{TableI}
\end{table*}

%%%%%%%%%%%%%%%%%%%%%%%%%%%%%%%%%%%%%%%%%%%%%%%%%%%%%%%%%%%%%%%%%%%%

This subsection applies the D3V55 model to analyze the 2D Riemann problem in gas dynamics. The 2D Riemann problem in gas dynamics exhibits rich and complex mesoscale structures\cite{schulz1993SJOSC,kurganov2002NMFPDE,lax1998SJOSC}. These structures, formed by interactions among shock waves, rarefaction waves, and contact discontinuities, result in 19 distinct configurations that challenge the stability and validity of the model.
High-order models typically exhibit reduced stability but higher accuracy. The successful simulation confirms that the D3V55 model combines robustness with high accuracy.

This study analyzes six selected configurations, with their initial conditions provide in Table \ref{TableI}. The symbols $(P_i, \rho_i, u_{x i}, u_{y i})$ represent the physical quantities in the $i$th quadrant, with $u_z=0$ across all quadrants. The simulation employs a grid of $N_x \times N_y \times N_z = 200 \times 200 \times 4$, applying Dirichlet boundary conditions in the $x$ direction and periodic conditions in the $y$ and $z$ directions. The grid spacing and time step are set as $\Delta x = \Delta y = \Delta z = 10^{-3}$, $\Delta t = 10^{-4}$, and the relaxation time as $\tau = 5 \times 10^{-5}$.

Figure \ref{Fig06} displays the density distribution for the corresponding configuration computed using the D3V55 model. The density contour lines clearly depict the interaction of various wave structures, such as shock waves, expansion waves, and contact discontinuities. These characteristic structures exhibit smooth and continuous profiles with high resolution, with no significant overshoot or spurious numerical oscillations observed. This demonstrates the model's ability to capture flow details and its robustness in managing complex wave configurations. %Furthermore, the model accurately captures the position and shape of the slip line, aiding the understanding of the rotation and vortex dynamics of the fluid in the 2D Riemann problem.

Specifically, for configuration (a) at $t=0.05$, the parameters are set as $n = 2$, $c = 1.6$, and $\eta_0 = 2.2$. The initial state consists of two shock waves and two contact discontinuities.
The slip lines $J_{32}$ and $J_{34}$ intersect the sound speed circle of the steady state in the third quadrant, creating a smooth transition at the interface. Here, $J_{ij}$ denotes the contact wave between the $i$th and $j$th quadrants [see Fig. \ref{Fig06}(a)]. The interaction between the shock waves generates a highly nonlinear flow field structure. Moreover, the figure clearly captures the triple-shock wave structure and the wave features in the first quadrant.

For configuration (b) at $t=0.05$, the parameters are $\tau = 9 \times 10^{-4}$, $n = 1$, $c = 1.6$, $\eta_0 = 1.8$. The interaction between shock waves $S_{21}$ and $S_{32}$ produces a triple-shock configuration and an elliptical subsonic region. The central density distribution displays a feather-like structure, with multiple rotating branches. This structure typically arises from fluid entering or exiting the computational domain or may result from some instability [see Fig. \ref{Fig06}(b)].

For configuration (c) at $t=0.05$, the parameters are $\tau = 10^{-5}$, $n = 2$, $c = 1.5$, $\eta_0 = 2.0$. The slip lines $J_{32}$ and $J_{34}$ in the third quadrant resemble those in Fig. \ref{Fig06} (a), with notable density gradients in the lower left and upper right corners, and multiple density jump regions in the center, indicating complex interactions or instability [see Fig. \ref{Fig06}(c)].

For configuration (d) at $t = 0.05$, the parameters are $\tau = 8 \times 10^{-5}$, $n = 1$, $c = 1.2$, $\eta_0 = 2.3$. Two shock waves ($S_{21}$ and $S_{41}$), two slip lines ($J_{32}$ and $J_{34}$), an elliptical subsonic region, and ripples in the first quadrant are clearly observed. Figure \ref{Fig06} (d) shows a more symmetric rotational flow field, with strong flow concentration in the central region.

For configuration (e) at $t = 0.05$, the parameters are $\tau = 10^{-5}$, $n = 1$, $c = 1.1$, $\eta_0 = 2.0$. The entire region is roughly divided into two parts by the slip lines $J_{21}$ and $J_{34}$, with a vortex in the subsonic region. The lower left corner shows a more prominent vortex structure, with higher density on the left side and lower density on the right, displaying strong asymmetry [see Fig. \ref{Fig06}(e)].

For configuration (f) at $t = 0.03$, the parameters are $\tau = 9 \times 10^{-5}$, $n = 2$, $c = 1.1$, $\eta_0 = 2.7$. A symmetric clockwise vortex forms, consisting of four slip lines. The density in the central region is very low, while the density in the outer region increases gradually, demonstrating a significant rotational pattern [see Fig. \ref{Fig06}(f)].

In summary, these results not only highlight the challenges posed by the complex mesoscale structures in the 2D gas dynamics Riemann problem, but also underscore the effectiveness of the DBM model in capturing intricate details. % that were previously unattainable with lower-order models.

\subsection{3D Riemann problem: spherical shock expansion in open space}
\begin{figure*}[htbp]
	{\centering
		\includegraphics[width=1.0\textwidth]{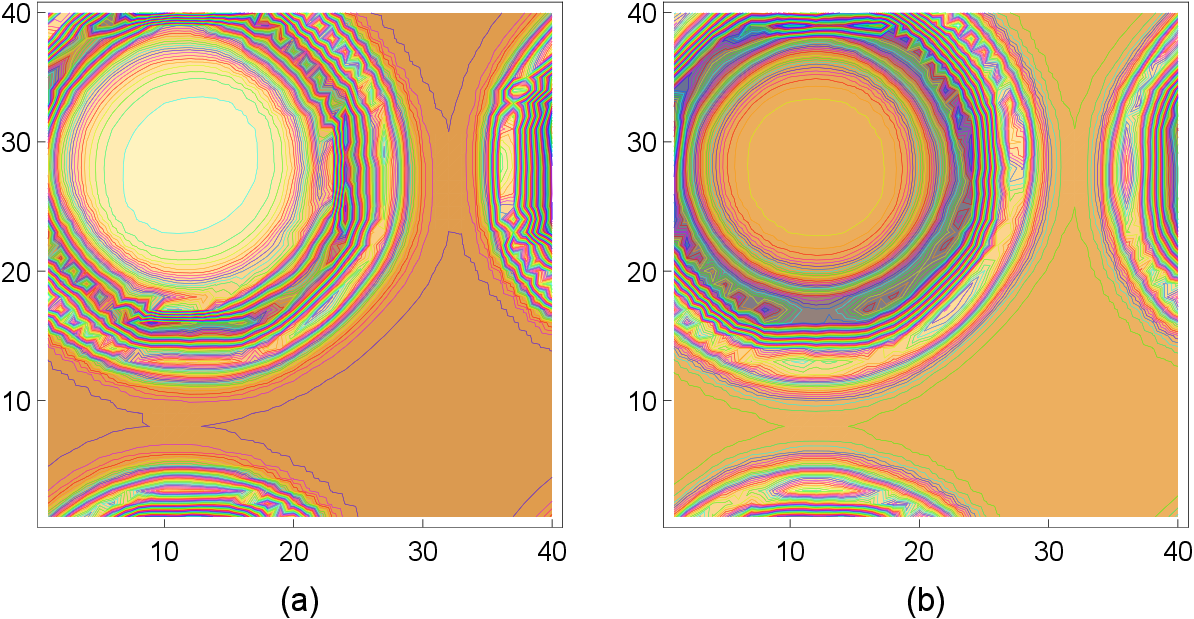}}
	\caption{\centering{Density distribution map of the 3D Riemann problem at $i_z=N_z / 2$ calculated using the D3V30 model (a), and the D3V55 model (b).}}
	\label{Fig07}
\end{figure*}

The spherical shock expansion problem is a classical issue in gas dynamics and nonlinear wave theory. It focuses on understanding the nonlinear propagation of shock waves, the evolution of wave system structures, and the interaction between the shock waves and the surrounding fluid. This problem is crucial for explosion safety design. Predicting shock wave propagation characteristics allows the assessment of potential hazards to the environment and personnel, enabling the development of protective plans and optimizing explosion protection structures, such as blast walls and protective shells.

This subsection simulates the spherical shock expansion problem in open space using both the NS-level 3D D3V30 model and the Burnett-level D3V55 model.
Simulating 3D problems imposes higher demands on the model than 1D and 2D problems, particularly regarding computational efficiency, stability, accuracy, boundary conditions, and nonequilibrium effects.
The simulation uses a grid of $N_x \times N_y \times N_z = 40 \times 40 \times 40$ with periodic boundary conditions in the $x$, $y$, and $z$ directions.
The initial velocity is set to zero, and the following conditions are applied: when $\sqrt{(i_x - 0.3 N_x)^2 + (i_y - 0.7 N_y)^2 + (i_z - 0.5 N_z)^2} \leq 0.3 N_x$, $\rho = 5.0$, $P = 5.0$; otherwise, $\rho = 1.0$, $P = 1.0$. Here, $i_x$, $i_y$, and $i_z$ represent the indices of the grid points.
The parameters for both the D3V30 and D3V55 models are: $\Delta x = \Delta y = \Delta z = 3 \times 10^{-3}$, $\Delta t = 5 \times 10^{-5}$, and $\tau = 5 \times 10^{-4}$. For the D3V30 model, $n = 2$, $c = 1.4$, and $\eta_0 = 2.0$. For the D3V55 model, $n = 2$, $c = 1.5$, and $\eta_0 = 2.2$. Figure \ref{Fig07} shows the density distribution at $i_z = N_z / 2$, with panel (a) and panel (b) are obtained from the D3V30 and D3V55 models, respectively. The results from both models show good agreement.

The simulation results clearly illustrate the nonlinear propagation effects in spherical shock expansion, including rapid density decay outward from the shock center and backward propagation of rarefaction waves. Initially, the high-pressure gas expands rapidly, compressing the surrounding low-pressure gas and forming the spherical shock. During expansion, nonlinear density and pressure variations create steep gradients in the shock front region, as shown by the densely packed contour lines.
As the shock propagates, the energy spreads over a larger spatial region, decreasing shock strength (density and pressure peaks), as reflected in the sparsity of the contour lines.
Behind the compression wave, high-density gas expansion generates rarefaction waves, which move opposite to the shock and interact with the fluid near the sphere's center.
Following the shock's outward expansion, the pressure difference across the shock front causes fluid to backfill, forming vortices and vortex structures. The contour lines in Figs. \ref{Fig07} (a) and (b) show disturbances in the post-shock region, with complex wave patterns emerging between the rarefaction wave and the main shock.

The D3V55 model, incorporating higher-order TNEs and nonlinear constitutive relations, provides a more precise representation of the distribution function's temporal evolution. This allows it to more effectively capture large gradients and strong nonlinear structures. Moreover, the model's ability to preserve the symmetry of spherical waves and handle the complexity of wave system evolution highlights its robustness in addressing high-dimensional hydrodynamic problems.

\section{Capability to capture higher-order thermodynamic nonequilibrium effects}\label{TNE}

The previous examples primarily demonstrate the model's ability to capture large-scale flow structures, focusing on the evolution of macroscopic, conserved, and slow varying variables. In contrast, this section evaluates the model's ability to represent mesoscopic, nonconserved, and fast varying variables.
In this study, the intensity of TNEs, as defined by the analytical expressions in Eqs. (\ref{D2})-(\ref{D31}) and Table \ref{TableI}, depends on macroscopic quantities, their gradients, relaxation times, and other factors. Thus, an intensity vector $\mathbf{S}_\text{TNE} = (\tau, \rho, T, \mathbf{u}, \bm{\nabla} \rho, \bm{\nabla} T, \bm{\nabla} \mathbf{u}, \bm{\Delta}_2^*, \bm{\Delta}_{3,1}^*)$, comprising nine elements, is defined to fully characterize the nonequilibrium manifestations.

The nonequilibrium intensity vector offers a multi-perspective description of nonequilibrium states and behaviors. Its components consist of characteristic quantities and key quantities of interest. Characteristic quantities are defined by the system's nonequilibrium driving mechanisms, while quantities of interest are determined by the system's characteristics and are highlighted in both traditional computational fluid dynamics and the DBM framework. Additional components can be incorporated into $\mathbf{S}_\text{TNE}$ as necessary to maintain the vector’s completeness.

Viscous stress and heat flux are key quantities in traditional fluid models. Adjusting elements in $\mathbf{S}_\text{TNE}$ generates TNEs of varying orders, enabling the exploration of the model's performance in cross-scale phenomena. To achieve this, a series of fluid collision experiments were conducted, adjusting initial conditions and physical parameters. The initial conditions are:
\begin{equation}\label{e44}
	\rho(x, y, z)=\frac{\rho_L+\rho_R}{2}-\frac{\rho_L-\rho_R}{2} \tanh \left(\frac{x-N_x \Delta x / 2}{L_\rho}\right),
\end{equation}
\begin{equation}\label{e45}
	u_x(x, y, z)=-u_0 \tanh \left(\frac{x-N_x \Delta x / 2}{L_u}\right),
\end{equation}
where \( L_\rho \) and \( L_u \) represent the widths of the density and velocity transition layers, respectively, and \( \rho_L \) and \( \rho_R \) are the densities outside the left and right fluid interfaces. The computational domain is a rectangular prism with dimensions \( 1.5 \times 0.006 \times 0.006 \), divided into a uniform grid of \( 1000 \times 4 \times 4 \).

\subsection{Typical thermodynamic nonequilibrium quantities: Viscous stress}

\begin{figure*}[htbp]
	{\centering
		\includegraphics[width=1.0\textwidth]{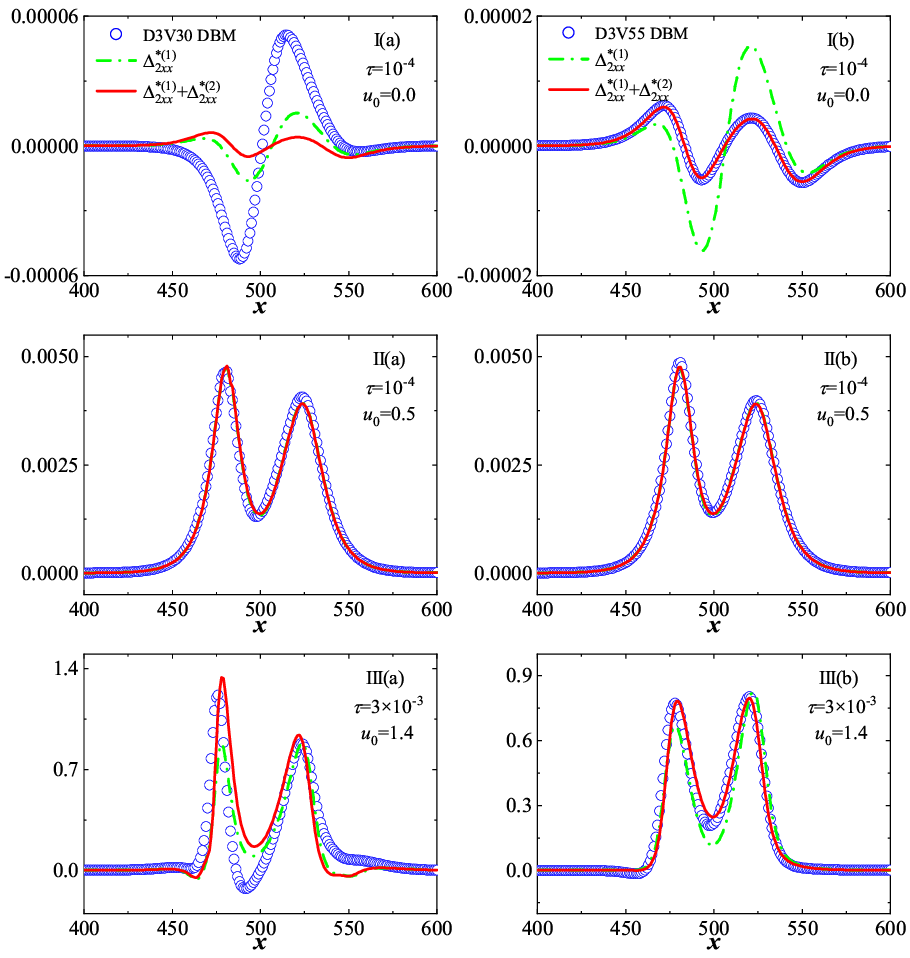}}
	\caption{\centering{Viscous stress calculated using the D3V30 model (left column) and the D3V55 model (right column) for three cases: weak (I), moderate (II), and strong (III). The green dashed line indicates the first-order analytical solution, while the red solid line represents the second-order analytical solution.}}
	\label{Fig08}
\end{figure*}

The initial values of other physical quantities are set as $\rho_L=2 \rho_R=2, P_L=2,P_R=1, u_y=0, u_z=0, L_\rho=L_u=20$.
By varying \( \tau \) and \( u_0 \), three different cases of viscous stress are examined: (i) \( \tau = 10^{-4}, u_0 = 0 \); (ii) \( \tau = 10^{-4}, u_0 = 0.5 \); and (iii) \( \tau = 4 \times 10^{-3}, u_0 = 1.5 \). Figure \ref{Fig08} shows the viscous stress distributions at \( t = 0.0025 \), calculated using both the NS-level D3V30 model and the Burnett-level D3V55 model. The NS-level model satisfies seven kinetic moment relations [Eqs. (\ref{e12})-(\ref{e18})], whereas the Burnett-level model requires nine [Eqs. (\ref{e12})-(\ref{e20})]. In Fig. \ref{Fig08}, the green dotted line represents the first-order analytical solution, the red solid line the second-order analytical solution, and the blue hollow circles the DBM numerical solution.

In case (I), the initial velocity gradient is zero, yielding zero first-order viscous stress. However, the second-order viscous stress arises due to density and temperature gradients. At this stage, the second-order term, \( \Delta_{2xx}^{*(2)} \), is significantly larger than the first-order term, \( \Delta_{2xx}^{*(1)} \). Over time, as the density and temperature gradients induce a velocity gradient, \( \Delta_{2xx}^{*(1)} \) gradually surpasses \( \Delta_{2xx}^{*(2)} \). Moreover, the second-order viscous stress provides negative feedback to the overall stress, causing the TNE strength to weaken as the temperature and density gradients decrease. In this case, the TNEs are weak due to the small relaxation time, \( \tau \), and the static initial conditions.
However, as shown in Fig. \ref{Fig08} I(a), the D3V30 model's simulation results do not match either the first-order analytical solutions, \( \Delta_{2xx}^{*(1)} \), or the second-order solutions, \( \Delta_{2xx}^{*(1)} + \Delta_{2xx}^{*(2)} \). In contrast, the D3V55 model's simulation results closely align with \( \Delta_{2xx}^{*(1)} + \Delta_{2xx}^{*(2)} \). This discrepancy highlights that the D3V30 model does not satisfy the higher-order moment relations required to capture the second-order TNEs. Consequently, the Newtonian viscous stress law fails, and the nonlinear effects of TNE become evident.

For case (II), by increasing the collision speed to \( u_0 = 0.5 \), the TNEs are amplified to 200 times that of case (I). As shown in Fig. \ref{Fig08} II(a) and II(b), the larger velocity gradient  causes the TNEs to be dominated primarily by the first-order term, \( \Delta_{2xx}^{*(1)} \). Moreover, the first-order and second-order viscous stress analytical solutions coincide, suggesting that second-order TNEs can be neglected.
In cases dominated by first-order TNEs, both the D3V30 and D3V55 models accurately capture the viscous stress. However, the D3V30 model's numerical solution deviates slightly from the analytical solution, particularly at the peaks and valleys of the TNEs, at \( x = 481 \), \( x = 500 \), and \( x = 525 \). This discrepancy arises because the linear constitutive relation in the lower-order model fails to correctly account for the contribution of velocity gradients to the second-order TNEs. Over time, the accumulation of constitutive errors leads to significant deviations in the macroscopic quantities computed with the D3V30 model compared to those from the higher-order model.

In case (III), as shown in Fig. \ref{Fig08} III(a) and III(b), increasing the relaxation time to \( \tau = 4 \times 10^{-3} \) and the collision speed to \( u_0 = 1.5 \) significantly enhances the nonequilibrium intensity. The large collision speed not only amplifies the first-order nonequilibrium quantity \( \Delta_{2xx}^{*(1)} \), but also induces considerable changes in the density and temperature gradients, which strongly excite the second-order nonequilibrium quantity \( \Delta_{2xx}^{*(2)} \). Compared to the previous two cases, the TNE effect in this case is more complex, with a stronger interaction between \( \Delta_{2xx}^{*(1)} \) and \( \Delta_{2xx}^{*(2)} \).
In this strongly nonequilibrium scenario, the NS-level D3V30 model fails to effectively capture the viscous stress near the interface, with its numerical solution deviating significantly from both the first-order and second-order analytical solutions. In contrast, the Burnett-level D3V55 model accurately simulates these effects. Furthermore, unlike in case I(b), in most regions, the second-order TNEs provide positive feedback, further intensifying the viscous stress.

The Knudsen number is a key physical quantity that characterizes the level of nonequilibrium. It is calculated as \( Kn = \lambda/L \), where \( \lambda = c_s \tau \), \( c_s = \sqrt{\gamma R T} \) is the local sound speed, and \( L \) is the characteristic length scale, defined by the macroscopic gradient, i.e., \( L = \phi/|\bm{\nabla} \phi| \). The maximum Knudsen numbers for cases I, II, and III are 0.0011, 0.0016, and 0.088, respectively, all exceeding 0.001. This indicates that the NS model's applicable region for continuous flow has been surpassed, whereas the D3V55 model remains applicable in the transition flow regime. Although the maximum Knudsen number for case I is smaller than that for case II, the D3V30 model performs better in simulating case II. This suggests that the Knudsen number alone is insufficient for model selection.
To address this, Gan \textit{et al.} introduced a dimensionless parameter, \( R_\text{TNE} = \left| \Delta_{2xx}^{*(2)} / \Delta_{2xx}^{*(1)} \right| \)
, to quantify the relative strength of TNEs. This parameter assesses the dominance of second-order TNEs over first-order ones \cite{Gan2018PRE}. The maximum \( R_\text{TNE} \) values for the three cases are 0.75 (at \( x = 500 \)), 0.035 (at \( x = 497 \)), and 1.14 (at \( x = 499 \)), respectively. Although the Knudsen number for case I is smaller, its relative TNE intensity is high, indicating that the Burnett-level and higher-order DBM models are necessary.

Subsequently, they extended the relative nonequilibrium strength to a TNE vector \( \bm{R}_\text{TNE} = \left| \bm{\Delta}_{m,n}^{*(j+1)} / \bm{\Delta}_{m,n}^{*(j)} \right| \)
, where \( \bm{\Delta}_{m,n}^{*(j+1)} \) and \( \bm{\Delta}_{m,n}^{*(j)} \) represent the \( (j+1) \)th and \( j \)th order TNEs, respectively. The deviation \( \Delta = \Delta_{\text{DBM}} - \Delta_{\text{Exact}} \) between the numerical solutions from DBM simulations and their analytical solutions is used to assess the suitability of DBM for the current problem. The current-order DBM accurately describes the problem only when both \( R_\text{TNE} \) and \( \Delta \) are sufficiently small. Otherwise, a higher-order DBM model should be constructed to capture the higher-order TNEs \cite{Gan2022JOFM}.

As the intensity of flow nonequilibrium increases, the distribution function of the fluid system deviates further from equilibrium, resulting in more complex variations in viscous stress. As a result, the demand for greater accuracy in viscous stress calculations increases. In high-speed compressible flows, traditional NS equations often fail to accurately describe mesoscopic structures, such as shock waves, boundary layers, and rarefied waves, where significant viscous stress occurs. Burnett-level and higher models, by reconstructing higher-order distribution functions, can more accurately capture the TNEs, significantly improving viscous stress calculation accuracy.

\begin{figure*}[htbp]
	{\centering
		\includegraphics[width=0.8\textwidth]{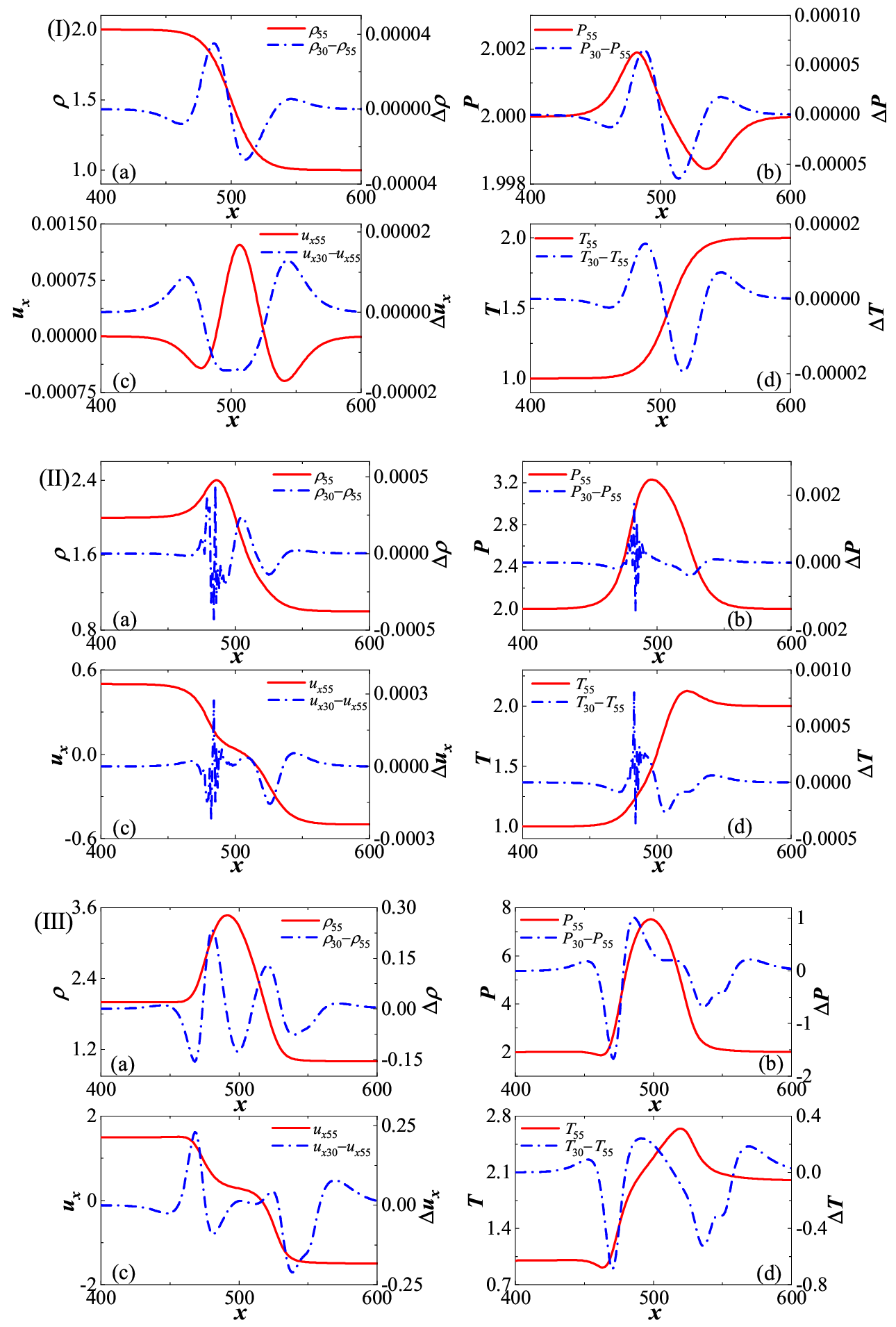}}
	\caption{\centering{Comparison of macroscopic quantities from the D3V55 model and their differences with the D3V30 model for cases I (top), II (middle), and III (bottom): (a) Density, (b) Pressure, (c) Velocity, (d) Temperature.}}
	\label{Fig09}
\end{figure*}

The accumulation of errors in describing viscous stress ultimately impacts the accuracy of macroscopic quantity calculations. Figure \ref{Fig09} illustrates that the red solid line represents macroscopic quantities computed using the D3V55 model, while the blue dashed line shows the differences in macroscopic quantities between the D3V30 and D3V55 models for various viscous stress scenarios. The differences in calculated macroscopic quantities for cases I, II, and III are $9\%$, $2\%$, and $66\%$ at \(t = 0.0025\), respectively, compared to the exact solution. The following conclusions can be drawn:

(i) The differences in macroscopic quantities (\(\Delta \rho\), \(\Delta P\), \(\Delta u_x\), and \(\Delta T\)) between the high-order and low-order models are closely linked to the intensity of nonequilibrium. In regions with large viscous stress differences, such as near local extremes (e.g., around \(x = 475\) and \(x = 525\)), the differences in macroscopic quantities also reach their peak values. As we move further from the interface, both the viscous stress and macroscopic quantity differences diminish.

(ii) The differences in macroscopic quantities between the high-order and low-order models are correlated with the relative strength of nonequilibrium. As the relative nonequilibrium strength increases (0.75, 0.035, and 1.14), the discrepancies between the models for each macroscopic quantity grow significantly. This suggests that second-order TNEs play a substantial role in system evolution, highlighting the necessity for higher-order physical models. Specifically, low-order nonequilibrium effects, represented by $\bm{\Delta}_2^*$ and $\bm{\Delta}_{3,1}^*$, function as constitutive relations and are embedded in the macroscopic equations, thereby influencing the evolution of macroscopic variables under different dissipation mechanisms. High-order nonequilibrium effects, represented by $\bm{\Delta}_3^*$ and $\bm{\Delta}_{4,2}^*$, serve as constitutive relations for $\bm{\Delta}_2^*$ and $\bm{\Delta}_{3,1}^*$, thereby governing the evolution equations of lower-order constitutive relations\cite{Gan2022JOFM}.

(iii) The viscous stress description errors from low-order models create a complex feedback loop with macroscopic quantity differences. For instance, near \(x = 475\), the low-order model overestimates the viscous stress (\(\Delta u_x > 0\)), while near \(x = 525\), it underestimates it (\(\Delta u_x < 0\)). In contrast, the feedback loop for \(\rho\), \(P\), and \(T\) is reversed. This occurs because the velocity gradient primarily contributes to first-order viscous stress, while \(\rho\), \(P\), and \(T\) contribute to second-order viscous stress.

\begin{figure*}[htbp]
	{\centering
		\includegraphics[width=0.92\textwidth]{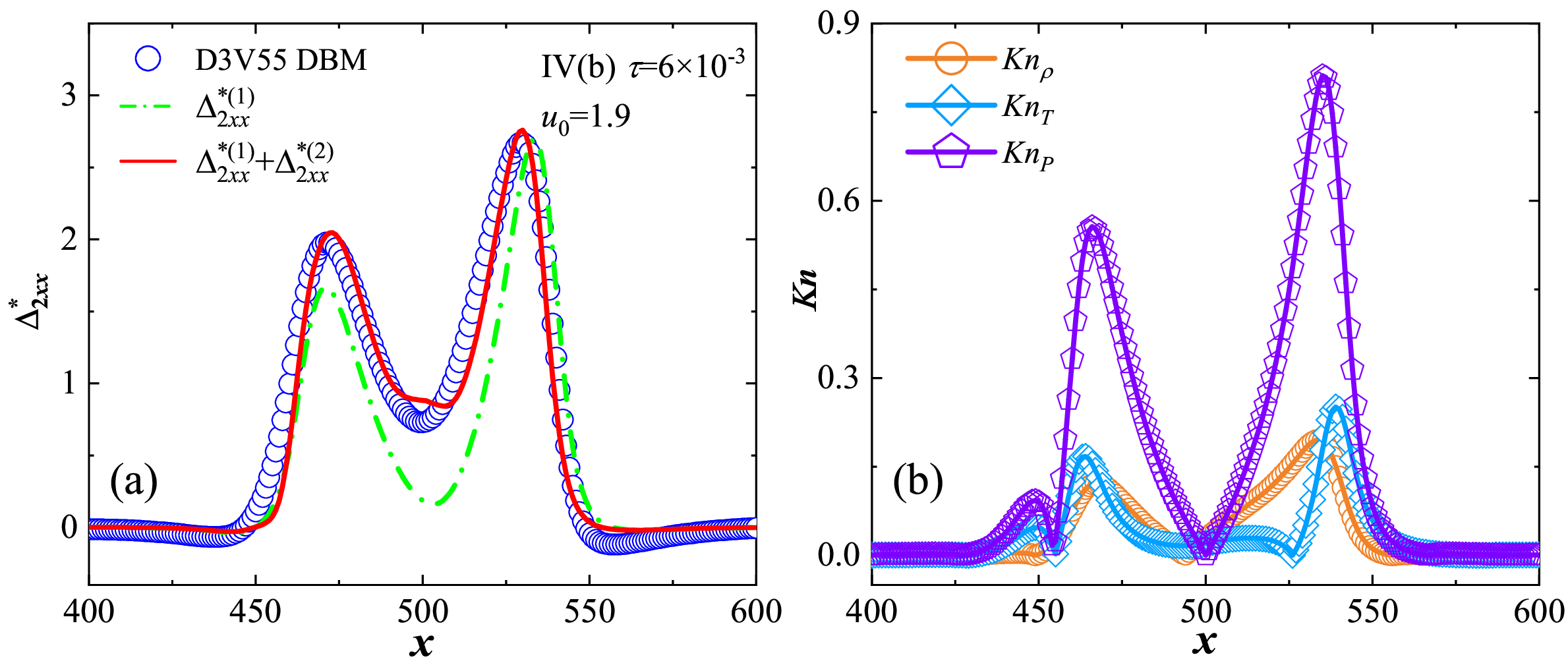}}
	\caption{\centering{Viscous stress at high nonequilibrium intensity (a) and the local Knudsen number calculated from pressure, density, and temperature (b).}}
	\label{Fig10}
\end{figure*}

To further assess the reliability of the D3V55 model in capturing stronger TNEs, \(\tau\) was increased to \(6 \times 10^{-3}\) and \(u_0\) was raised to 1.9. At this point, \(R_{\text{TNE}}\) reached a value of 4.39 (at \(x = 503\)). As shown in Fig. \ref{Fig10} (a), \(\Delta_{2xx}^{*(1)}\) failed to accurately capture the peak region, while the DBM numerical solution matches \(\Delta_{2xx}^{*(1)} + \Delta_{2xx}^{*(2)}\) well. This indicates that the second-order stress dominates in high nonequilibrium regions, particularly where velocity and temperature gradients vary drastically. The D3V55 model clearly exhibits a stronger ability to describe TNEs compared to the 2D Burnett-level DBM.
Figure \ref{Fig10} (b) shows the local Knudsen number calculated from density, temperature, and pressure. Its distribution along the \(x\)-axis is similar, but not identical, to the distribution of viscous stress, indicating that different macroscopic quantities have distinct characteristic structures. This is because (i) their dominant and dissipation mechanisms differ, and (ii) the Knudsen number, calculated from the gradient of conserved quantities, is a slowly varying and simpler variable, whereas nonequilibrium quantities are fast-varying, nonlinear combinations of macroscopic variables and their gradients, which more accurately reflect the system’s deviation from equilibrium and corresponding effects. As a typical nonequilibrium quantity, the distribution of viscous stress depends on the distributions of all macroscopic quantities and their nonlinear coupling. The maximum Knudsen numbers calculated from density and temperature are 0.20 and 0.25, respectively, while the Knudsen number calculated from pressure reaches 0.81, indicating that the flow has entered the transition regime.

\begin{figure}[htbp]
	{\centering
		\includegraphics[width=0.49\textwidth]{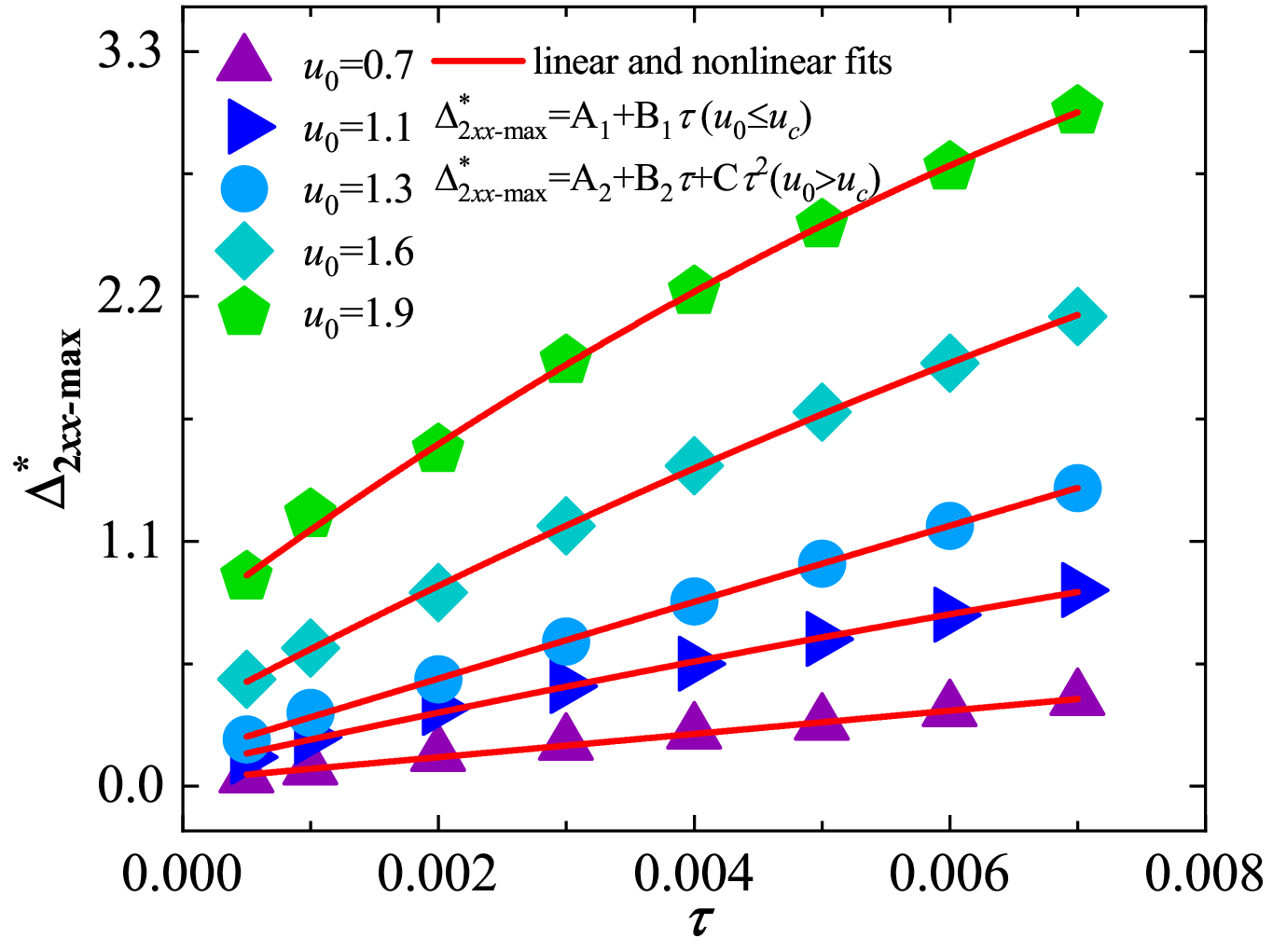}}
	\caption{\centering{Linear and nonlinear viscous stress.}}
	\label{Fig11}
\end{figure}

Further investigations were conducted to examine the effects of velocity and relaxation time on viscous stress. Figure \ref{Fig11} illustrates how the maximum viscous stress \( \Delta_{2xx-\text{max}}^* \) varies with relaxation time \( \tau \) for different initial velocities \( u_0 \). The red solid lines represent the results of both linear and nonlinear fitting. The figure clearly shows that as relaxation time and flow velocity increase, the strength of the TNEs also grows. Additionally, a critical flow velocity \( u_c (= 1.6) \) exists: when \( u_0 \) is below this threshold, the maximum viscous stress \( \Delta_{2xx-\text{max}}^* \) increases linearly with relaxation time \( \tau \), as described by the relation \( \Delta_{2xx-\text{max}}^* = A_1 + B_1 \tau \). However, when \( u_0 \) exceeds this threshold, the maximum viscous stress increases nonlinearly with relaxation time, in accordance with the relation \( \Delta_{2xx-\text{max}}^* = A_2 + B_2 \tau + C \tau^2 \). This indicates that, under strong nonequilibrium conditions, the linear constitutive relations used in the NS fluid model break down, and the physical model must account for higher-order deviations \( f^j (j \geq 2) \) from equilibrium in the distribution function. Similarly, the flow velocity has a comparable effect on other TNE quantities. In fact, the magnitude of the relative nonequilibrium strength determines the transition between linear and nonlinear behavior.

\begin{figure}[htbp]
	{\centering
		\includegraphics[width=0.45\textwidth]{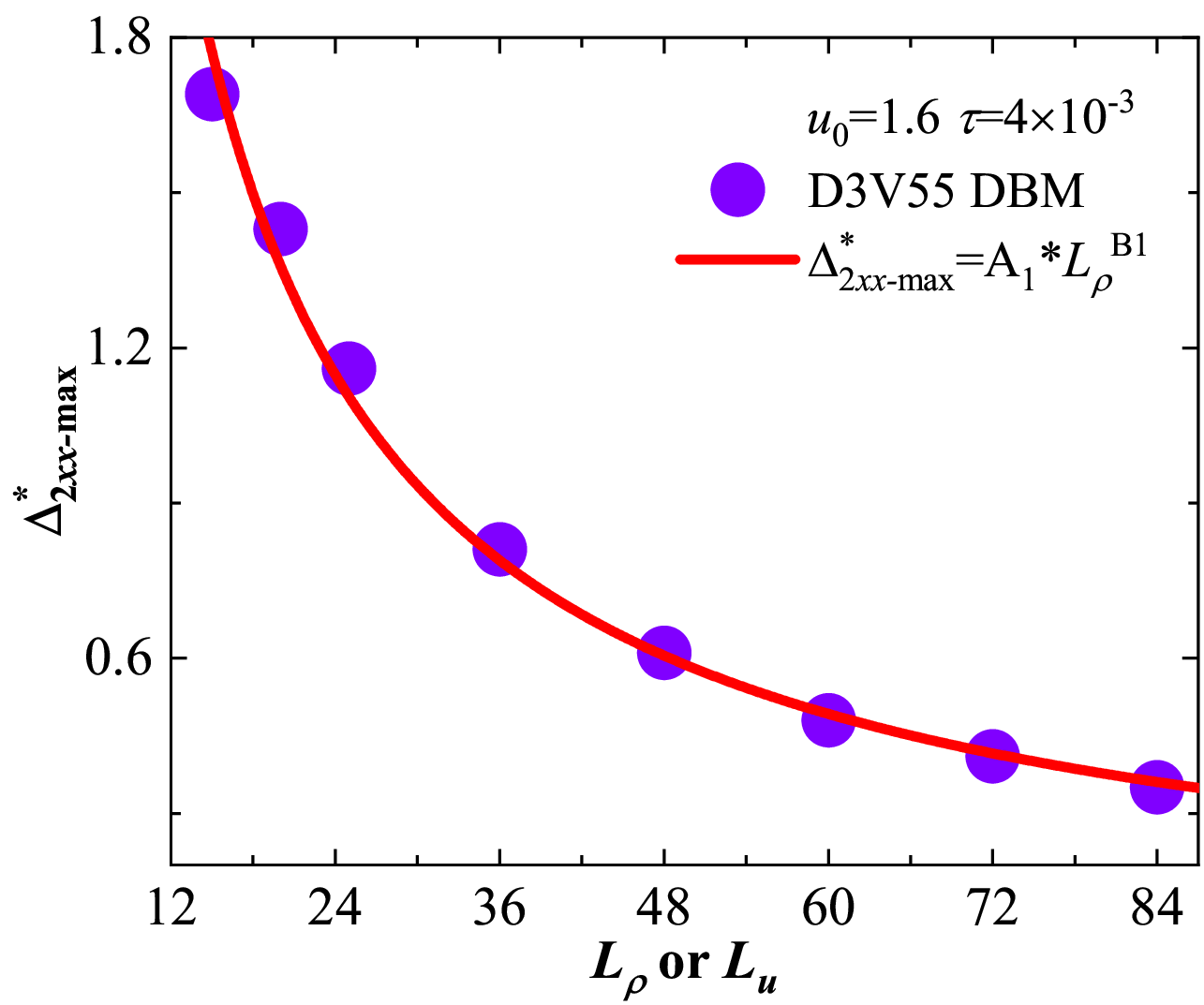}}
	\caption{\centering{Effects of interface width on viscous stress.}}
	\label{Fig12}
\end{figure}

The interface widths \(L_\rho\) and \(L_u\) also affect the TNEs. With \(u_0 = 1.6\) and \(\tau = 4 \times 10^{-3}\), Figure \ref{Fig12} demonstrates that as the interface widths \(L_\rho\) and \(L_u\) increase, the maximum viscous stress \(\Delta_{2xx-\text{max}}^*\) gradually decreases. This occurs because increasing the interface width reduces the macroscopic gradient, thereby suppressing the strength of the TNEs. The relationship between \(\Delta_{2xx-\text{max}}^*\) and \(L_\rho\) (or \(L_u\)) follows a negative power-law:
$\Delta_{2xx-\text{max}}^* = A_1 \cdot L_\rho^{B_1}$, where \(A_1 \approx 21.65\) and \(B_1 \approx -0.92\). The curve fitting has a correlation coefficient  \(R^2 = 0.98\), indicating the goodness of the fit.

\subsection{Typical thermodynamic nonequilibrium quantity: Heat flux}

\begin{figure*}[htbp]
	{\centering
		\includegraphics[width=0.92\textwidth]{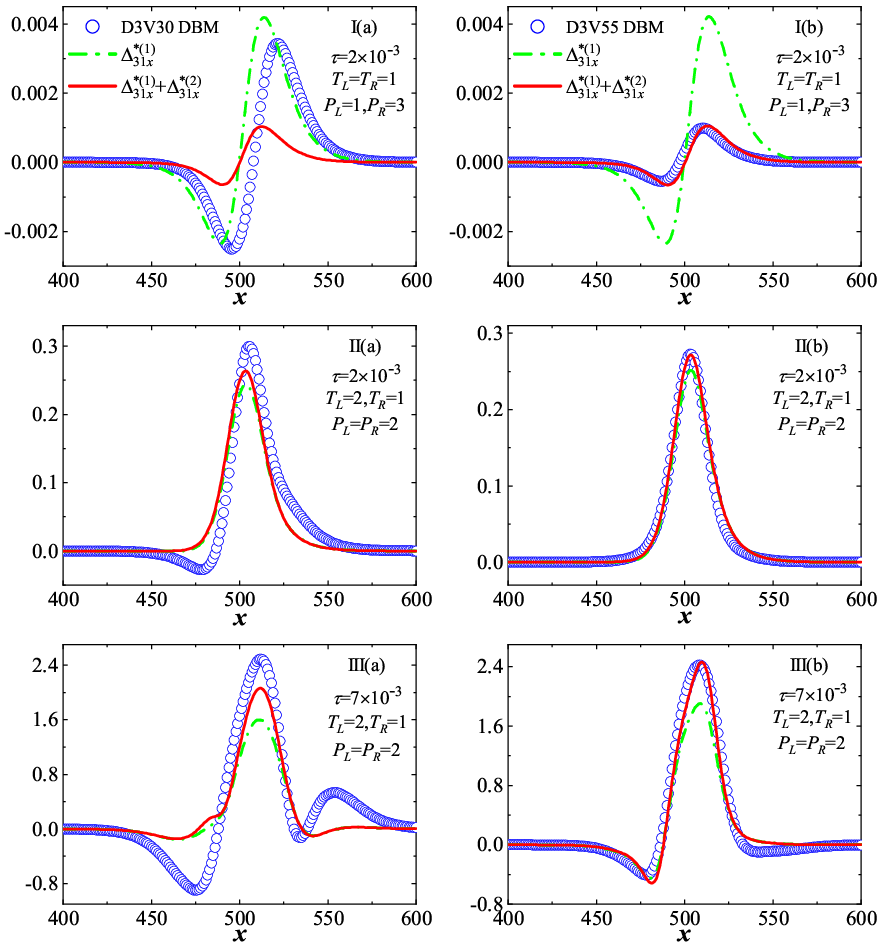}}
	\caption{\centering{ Heat flux calculated using the D3V30 model (left column) and the D3V55 model (right column) for three cases: weak (I), moderate (II), and strong (III). The green dashed line indicates the first-order analytical solution, while the red solid line represents the second-order analytical solution.}}
	\label{Fig13}
\end{figure*}

Similarly, the ability of the D3V55 model to capture another typical nonequilibrium quantity, heat flux, is examined. The initial conditions and parameters for the three cases are set as follows: (i) \( T_L = T_R = 1, P_L = 1, P_R = 3, u_0 = 0.5, \tau = 2 \times 10^{-3} \); (ii) \( T_L = 2, T_R = 1, P_L = P_R = 2, u_0 = 1.0, \tau = 2 \times 10^{-3} \); (iii) \( u_0 = 1.4, \tau = 7 \times 10^{-3} \). Figure \ref{Fig13} shows the heat flux computed using the D3V30 model at the NS level (left column) and the D3V55 model at the Burnett level (right column) for three cases. The results for cases I, II, and III are at \( t = 0.0005, 0.007, \) and \( 0.00175 \), respectively. The green dashed line represents the first-order analytical solution, the red solid line represents the second-order analytical solution, and the blue hollow circle represents the numerical solution.

In case I, the initial temperature is uniform, resulting in a nearly zero first-order heat flux, \( \Delta_{3,1x}^{*(1)} \), as the system evolves. The second-order heat flux, \( \Delta_{3,1x}^{*(2)} \), is initially triggered by the density gradient, and is further excited by the temperature and velocity gradients, which lead to the emergence of the first-order heat flux. As shown in Fig. \ref{Fig13} I(a), when the second-order TNEs dominate, the heat flux computed by the D3V30 model deviates significantly from the analytical solution. In contrast, the D3V55 model’s numerical solution closely matches the second-order analytical solution, \( \Delta_{3,1x}^{*(1)} + \Delta_{3,1x}^{*(2)} \), emphasizing the necessity of accounting for second-order TNEs.

\begin{figure*}[htbp]
	{\centering
		\includegraphics[width=0.92\textwidth]{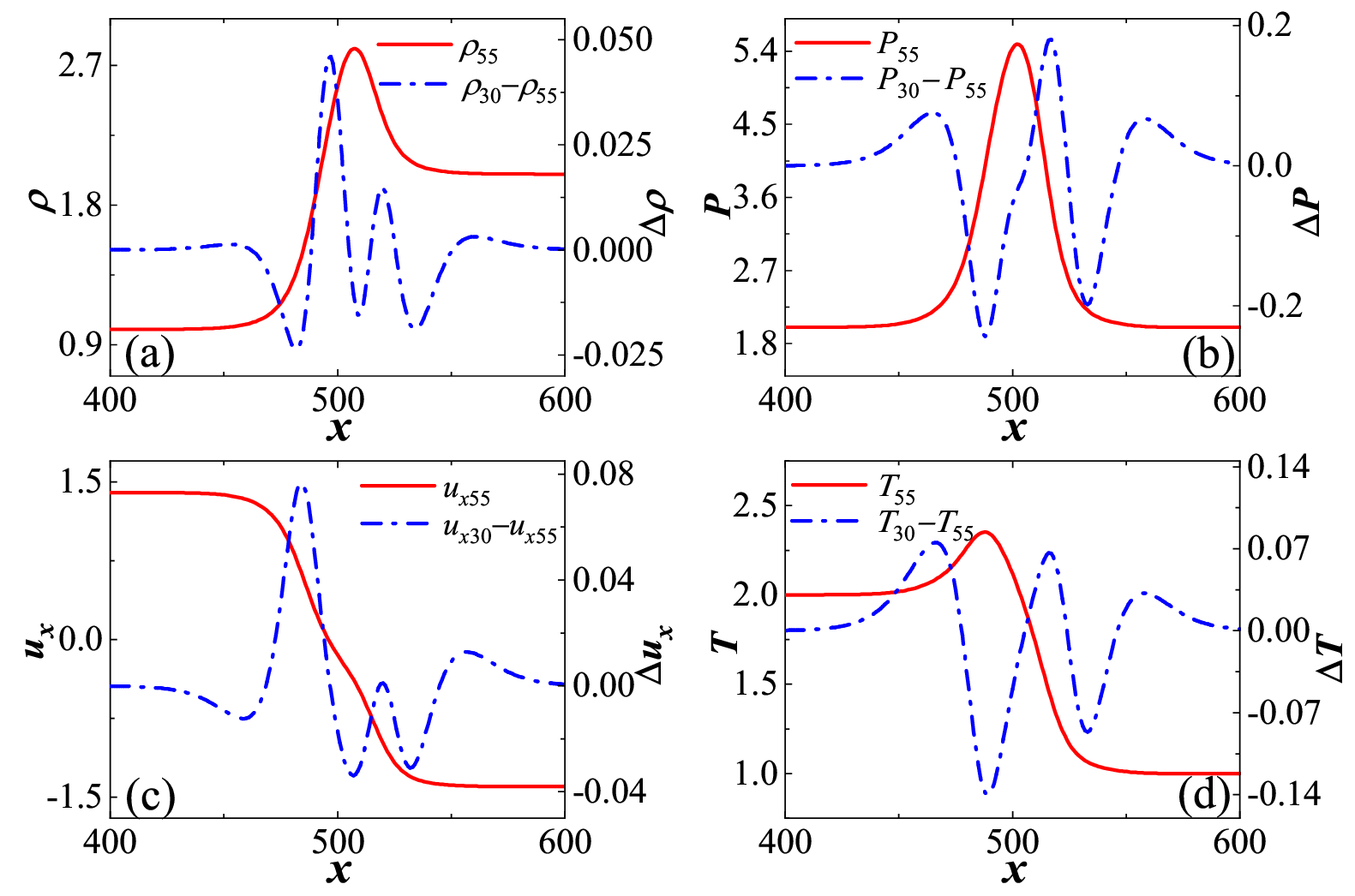}}
	\caption{\centering{ Comparison of macroscopic quantities from the D3V55 model and their differences with the D3V30 model for case III: (a) Density, (b) Pressure, (c) Velocity, (d) Temperature.}}
	\label{Fig14}
\end{figure*}

In case II, increasing the temperature gradient and velocity significantly enhances the amplitude of the heat flux, contributed by both the first-order and second-order TNEs. However, in this case, \( R_\text{TNE} = 0.08 \), and the second-order nonequilibrium intensity remains relatively weak. Nonetheless, the D3V30 model fails to accurately describe the system's heat flux, with both its amplitude and distribution deviating significantly from the analytical solution. As shown in Fig. \ref{Fig13} II(b), the D3V55 model more accurately captures the second-order TNEs, with its numerical solution closely matching the second-order analytical solution. The second-order quantity, \( \Delta_{3,1x}^{*(2)} \), provides weak positive feedback to the overall nonequilibrium.

In case III, the relaxation time, \( \tau \), is further increased to \( 7 \times 10^{-3} \), and the collision velocity, \( u_0 \), is raised to 1.4, leading to an increased heat flux. In Fig. \ref{Fig13} III(b), the numerical solution closely matches the second-order analytical solution. Unlike case I, the second-order nonequilibrium significantly amplifies the overall nonequilibrium. As the nonequilibrium strength of the fluid system increases, accurate heat flux calculations become crucial for describing the system. For example, in the shockwave region, molecular collisions and transport effects intensify heat flux fluctuations, necessitating higher-order models to capture these microscopic phenomena accurately. Therefore, Burnett-level and higher models are indispensable.

Figure \ref{Fig14} presents a comparison of macroscopic quantities computed using the D3V55 model with those from the D3V30 model for case III at \( t = 0.00175 \). The differences arise from variations in their ability to describe thermodynamic nonequilibrium effects. Some nonequilibrium quantities are directly embedded into the macroscopic fluid equations, determining the evolution and dissipation of macroscopic quantities. The computed difference in macroscopic quantities for case III is $22\%$, indicating that, in strong nonequilibrium situations, the distribution of macroscopic quantities deviates from the continuous medium assumption, exhibiting significant discretization. In Burnett-level heat flux cases, the differences between the two models in each macroscopic quantity gradually increase with the accumulation of relative nonequilibrium strength, \( R_{\text{TNE}} \), over time. These differences, due to the physical model's accuracy, will not diminish with improvements in numerical precision.

\section{Conclusions, discussions and outlooks} \label{Conclusions} %and outlooks

Three-dimensional (3D) high-speed compressible flows represent a complex interplay of nonlinear, nonequilibrium, and multiscale phenomena. These flows present formidable challenges to traditional fluid models, particularly in capturing the significant discrete effects and intricate thermodynamic nonequilibrium effects (TNEs) that emerge under extreme conditions. To address these challenges, this study develops a discrete Boltzmann modeling and simulation method, which fuses the insights of kinetic theory with mean-field theory.
By using the Chapman-Enskog multiscale method, a set of kinetic moment relations
$\bm{\Phi}=(\mathbf{M}_0,\mathbf{M}_1,\mathbf{M}_{2,0},\mathbf{M}_2,\mathbf{M}_{3,1},\mathbf{M}_3,
\mathbf{M}_{4,2},\mathbf{M}_4,\mathbf{M}_{5,3})$ for describing second-order TNEs are identified.
These kinetic moment relations $\bm{\Phi}$ are crucial invariants that must be preserved in coarse-grained physical modeling, providing a fresh and deeper perspective for analyzing the TNE behaviors and their effects.
A discrete Boltzmann model, employing 55 discrete velocities to discretize the
3D velocity space, is proposed for multiscale simulations of 3D supersonic flows.
The nonlinear constitutive relations, as a by-product, are also derived,
contributing to the theoretical foundation of macroscopic fluid modeling.
Compared to the 2D case, the nonlinear constitutive relations in three dimensions are more complex due to the increased degrees of freedom, which lead to a greater variety of nonequilibrium driving forces, enhanced coupling between different nonequilibrium driving forces, and a significant increase in nonequilibrium components.
Extensive numerical validations demonstrate the model's accuracy across test cases ranging from 1D to 3D scenarios, and from subsonic to supersonic regimes.
Particularly, the model effectively captures multiscale TNEs, including viscous stress and heat flux, near mesoscale structures.

In principle, the extended Burnett equations (EBE) can also be used to study these mesoscale behaviors. However, deriving the full EBE is inherently difficult, and its high degree of nonlinearity and numerous nonlinear terms pose significant challenges for numerical simulations.
These combined complexities render the approach of ``first deriving the EBE, then numerically solving it'' almost impractical.
The DBM enables some of these studies, which become increasingly infeasible with higher Knudsen numbers, feasible.  Moreover, compared to the kinetic macro modeling (KMM) approach of ``first deriving the EBE, then solving it numerically'', the DBM offers new methodologies for analyzing the vast datasets generated from simulations. It particularly excels in answering the critical questions of ``how to describe discrete states and effects'', and ``how to extract and analyze nonequilibrium states and effects'' in a rigorous and efficient manner. In this study, we assess the model's applicability by evaluating the self-consistency of the numerical and analytical solutions for nonequilibrium effects, as well as the relative nonequilibrium intensity. The numerical results (Fig. \ref{Fig10}) show that when the Knudsen number computed from the density gradient reaches 0.2 (with Knudsen numbers of 0.25 and 0.81 calculated from the temperature and pressure gradients, respectively), the model proves physically effective and is suitable for studying nonequilibrium problems across continuum to transition flow regimes.

Despite its strengths, the D3V55 DBM has certain limitations. First, it currently accounts only for second-order TNEs, excluding higher-order effects that become crucial as the Mach number increases and compressibility strengthens.
Second, the utilization of a single relaxation time simplifies modeling and computations, but fails to capture the varying rates at which different kinetic modes approach equilibrium, particularly when the disparities are prominent in complex or strongly nonequilibrium flows. Additionally, stability issues can also arise in some cases,  especially at high Knudsen and Mach numbers.

Future research will address these limitations in three key areas: (1) Incorporating higher-order TNEs, such as third-order effects, to improve accuracy under extreme nonequilibrium conditions; (2) To better account for the varying timescales of different kinetic modes, researchers have developed multi-relaxation-time higher-order models \cite{yan2024lattice,li2024self}; and (3) Conducting von Neumann stability analyses \cite{Gan2008PA} to identify factors influencing model stability and propose strategies to enhance its robustness and reliability across diverse flow conditions. These advancements are expected to broaden the model's applicability to aerospace engineering, energy extraction, security and defense, and other advanced technological fields.

\begin{acknowledgments}
The authors sincerely thank Zhaowen Zhuang, Jiahui Song for helpful discussions.
We acknowledge support from the National Natural Science Foundation of China
(Grant Nos. U2242214, 52278119, 11875001 and 12172061), Hebei Outstanding Youth Science Foundation (Grant No. A2023409003), Central Guidance on Local Science and Technology Development Fund of Hebei Province (Grant No. 226Z7601G), Fujian Provincial Units Special Funds for Education and Research (Grant No. 2022639), the Foundation of National Key Laboratory of
Shock Wave and Detonation Physics (Grant no. JCKYS2023212003), the Opening Project of State Key Laboratory of Explosion Science and Safety Protection (Beijing Institute of Technology) (Grant No. KFJJ25-02M), and science foundation of NCIAE (Grant No. ZD-2025-06).
\end{acknowledgments}

\appendix
\section{Second-order viscous stress and heat flux in three dimensions}\label{App}

The expressions for the second-order viscous stress \( \bm{\Delta}_{2}^{*(2)} \) and heat flux \( \bm{\Delta}_{3,1}^{*(2)} \) in three-dimensions are presented in Table \ref{TableII}, with the associated terms \( S_1, \dots, S_{36} \) listed in Table \ref{TableIII}.

\begin{longtable}{|c|>{\raggedright\arraybackslash}p{5cm}|>{\raggedright\arraybackslash}p{3cm}|}
\caption{Expressions for  \( \bm{\Delta}_{2}^{*(2)} \) and \( \bm{\Delta}_{3,1}^{*(2)} \). } \label{TableII} \\ \hline
\textbf{TNE} & \textbf{Formula} & \textbf{Physical essence} \\ \hline
\( \Delta_{2xx}^{*(2)} \) &
\(-2   n_3^{-2} \tau^2  (n_3 R^2 T^2 S_1 - n_{-1} \rho R T S_2 +\rho R T S_3 - R^2 T^2 \frac{S_4}{\rho} - \rho R^2 S_5)\) & \multirow{6}{3cm}{non-organized momentum flux (NOMF)/generalized viscous stress} \\ \cline{1-2}
\( \Delta_{2xy}^{*(2)} \) &
\(-2  n_3^{-1} \tau^2 (\rho R T \partial_y u_x S_6 - \rho R T \partial_x u_y S_7 - \frac{n_3 R S_8}{\rho})\) & \\ \cline{1-2}
\( \Delta_{2xz}^{*(2)} \) &
\(-2 n_3^{-1} \tau^2 (\rho R T \partial_z u_x S_9 - \rho R T \partial_x u_z S_{10} - \frac{n_3 R S_{11}}{\rho})\) & \\ \cline{1-2}
\( \Delta_{2yy}^{*(2)} \) &
\(-2 n_3^{-2} R \tau^2(n_3 R T^2 S_{12} - n_{-1}\rho T S_{13} + \rho T S_{14} - R T^2 \frac{S_{15}}{\rho} - \rho R S_{16})\) & \\ \cline{1-2}
\( \Delta_{2yz}^{*(2)} \) &
\(-2 n_3^{-1} \tau^2 (\rho R T \partial_z u_y S_{17} - \rho R T \partial_y u_z S_{18} - \frac{n_3 R S_{19}}{\rho})\) & \\ \cline{1-2}
\( \Delta_{2zz}^{*(2)} \) &
\(-2 n_3^{-2} R \tau^2(n_3 R T^2  S_{20} - n_{-1} \rho T S_{21} + \rho T S_{22} - R T^2 \frac{S_{23}}{\rho} - \rho R S_{24})\) & \\ \hline
\( \Delta_{3,1x}^{*(2)} \) &
\(n_3^{-1} \tau^2 \rho R^2 T(T S_{25} + \partial_x T S_{26} + n_3 \partial_y T S_{27} + n_3 \partial_z T S_{28})\) & \multirow{3}{3cm}{non-organized energy flux (NOEF)/generalized heat flux} \\ \cline{1-2}
\( \Delta_{3,1y}^{*(2)} \) &
\(n_3^{-1} \tau^2 \rho R^2 T(T S_{29} + \partial_y T S_{30} + n_3 \partial_x T S_{31} + n_3 \partial_z T S_{32})\) & \\ \cline{1-2}
\( \Delta_{3,1z}^{*(2)} \) &
\(n_3^{-1} \tau^2 \rho R^2 T(T S_{33} + \partial_z T S_{34} + n_3 \partial_x T S_{35} + n_3 \partial_y T S_{36})\) & \\ \hline
\end{longtable}

\begin{longtable}{|c|p{7.5cm}|}
\caption{Expressions for $S_i$, $i=1,...,36$.} \label{TableIII}\\ \hline
\textbf{Index} & \textbf{Expression} \\ \hline
\( S_1 \) & \( n_2 \frac{\partial^2}{\partial x^2} \rho - \frac{\partial^2}{\partial y^2} \rho - \frac{\partial^2}{\partial z^2} \rho \) \\ \hline
\( S_2 \) & \( n_2 (\partial_x u_x)^2 - (\partial_y u_y)^2 - (\partial_z u_z)^2 \) \\ \hline
\( S_3 \) & \( 4 n_1 \partial_x u_x(\partial_y u_y+\partial_z u_z)-8 \partial_y u_y \partial_z u_z-n_2[(\partial_y u_x)^2+(\partial_z u_x)^2]+(\partial_x u_y)^2+(\partial_z u_y)^2+(\partial_x u_z)^2+(\partial_y u_z)^2 \) \\ \hline
\( S_4 \) & \( n_2 (\partial_x \rho)^2 + n_3 (\partial_y \rho)^2 + n_3 (\partial_z \rho)^2 \) \\ \hline
\( S_5 \) & \( n_2 (\partial_x T)^2 + n_3 (\partial_y T)^2 + n_3 (\partial_z T)^2 \) \\ \hline
\( S_6 \) & \( 2 \partial_x u_x - n_1 \partial_y u_y + \partial_z u_z \) \\ \hline
\( S_7 \) & \( n_1 \partial_x u_x - 2 \partial_y u_y - 2 \partial_z u_z \) \\ \hline
\( S_8 \) & \( \rho^2 T \partial_z u_x \partial_z u_y + R T^2 \partial_x \rho \partial_y \rho + \rho^2 R \partial_x T \partial_y T - \rho R T^2 \frac{\partial^2}{\partial x \partial y} \rho \) \\ \hline
\( S_9 \) & \( 2 \partial_x u_x + \partial_y u_y- n_1 \partial_z u_z  \) \\ \hline
\( S_{10} \) & \( n_1 \partial_x u_x - 2 \partial_y u_y - 2 \partial_z u_z \) \\ \hline
\( S_{11} \) & \( \rho^2 T \partial_y u_x \partial_y u_z + R T^2 \partial_x \rho \partial_z \rho + \rho^2 R \partial_x T \partial_z T - \rho R T^2 \frac{\partial^2}{\partial x \partial z} \rho \) \\ \hline
\( S_{12} \) & \( - \frac{\partial^2}{\partial x^2} \rho+n_2 \frac{\partial^2}{\partial y^2} \rho - \frac{\partial^2}{\partial z^2} \rho \) \\ \hline
\( S_{13} \) & \( - (\partial_x u_x)^2 + n_2 (\partial_y u_y)^2 - (\partial_z u_z)^2 \) \\ \hline
\( S_{14} \) & \( 4 n_1 \partial_y u_y(\partial_x u_x+\partial_z u_z)-8 \partial_x u_x \cdot \partial_z u_z-n_2[(\partial_x u_y)^2+(\partial_z u_y)^2]+(\partial_y u_x)^2+(\partial_z u_x)^2+(\partial_x u_z)^2+(\partial_y u_z)^2 \) \\ \hline
\( S_{15} \) & \( n_3 (\partial_x \rho)^2 +n_2 (\partial_y \rho)^2 + n_3 (\partial_z \rho)^2 \) \\ \hline
\( S_{16} \) & \( n_3 (\partial_x T)^2 + n_2 (\partial_y T)^2 + n_3 (\partial_z T)^2 \) \\ \hline
\( S_{17} \) & \( 2 \partial_x u_x + \partial_y u_y - n_1 \partial_z u_z \) \\ \hline
\( S_{18} \) & \( - 2 \partial_x u_x + n_1 \partial_y u_y - 2 \partial_z u_z \) \\ \hline
\( S_{19} \) & \( \rho^2 T \partial_x u_y \partial_x u_z + R T^2 \partial_y \rho \partial_z \rho + \rho^2 R \partial_y T \partial_z T - \rho R T^2 \frac{\partial^2}{\partial y \partial z} \rho \) \\ \hline
\( S_{20} \) & \(  - \frac{\partial^2}{\partial x^2} \rho - \frac{\partial^2}{\partial y^2} \rho + n_2 \frac{\partial^2}{\partial z^2} \rho \) \\ \hline
\( S_{21} \) & \( - (\partial_x u_x)^2 - (\partial_y u_y)^2 + n_2 (\partial_z u_z)^2 \) \\ \hline
\( S_{22} \) & \( 4 n_1 \partial_z u_z(\partial_x u_x+\partial_y u_y)-8 \partial_x u_x \cdot \partial_y u_y-n_2[(\partial_x u_z)^2+(\partial_y u_z)^2]+(\partial_y u_x)^2+(\partial_z u_x)^2+(\partial_x u_y)^2+(\partial_z u_y)^2 \) \\ \hline
\( S_{23} \) & \( n_3 (\partial_x \rho)^2 + n_3 (\partial_y \rho)^2 + n_2 (\partial_z \rho)^2 \) \\ \hline
\( S_{24} \) & \( n_3 (\partial_x T)^2 + n_3 (\partial_y T)^2 + n_2 (\partial_z T)^2 \) \\ \hline
\( S_{25} \) & \( n_{-1} \frac{\partial^2}{\partial x^2} u_x + n_3 \frac{\partial^2}{\partial y^2} u_x + n_3 \frac{\partial^2}{\partial z^2} u_x - 4 \frac{\partial^2}{\partial x \partial y} u_y - 4 \frac{\partial^2}{\partial x \partial z} u_z \) \\ \hline
\( S_{26} \) & \( (n^2 + 10 n + 13) \partial_x u_x - 2 n_7 (\partial_y u_y + \partial_z u_z) \) \\ \hline
\( S_{27} \) & \( n_7 \partial_y u_x + 2 \partial_x u_y \) \\ \hline
\( S_{28} \) & \( n_7 \partial_z u_x + 2 \partial_x u_z \) \\ \hline
\( S_{29} \) & \( n_3 \frac{\partial^2}{\partial x^2} u_y + n_{-1} \frac{\partial^2}{\partial y^2} u_y + n_3 \frac{\partial^2}{\partial z^2} u_y - 4 \frac{\partial^2}{\partial x \partial y} u_x - 4 \frac{\partial^2}{\partial y \partial z} u_z \) \\ \hline
\( S_{30} \) & \( (n^2 + 10 n + 13) \partial_y u_y - 2 n_7 (\partial_x u_x + \partial_z u_z) \) \\ \hline
\( S_{31} \) & \( 2 \partial_y u_x + n_7 \partial_x u_y \) \\ \hline
\( S_{32} \) & \( n_7 \partial_z u_y + 2 \partial_y u_z \) \\ \hline
\( S_{33} \) & \( n_3 \frac{\partial^2}{\partial x^2} u_z + n_3 \frac{\partial^2}{\partial y^2} u_z + n_{-1} \frac{\partial^2}{\partial z^2} u_z - 4 \frac{\partial^2}{\partial x \partial z} u_x - 4 \frac{\partial^2}{\partial y \partial z} u_y \) \\ \hline
\( S_{34} \) & \( (n^2 + 10 n + 13) \partial_z u_z - 2 n_7 (\partial_x u_x + \partial_y u_y) \) \\ \hline
\( S_{35} \) & \( 2 \partial_z u_x + n_7 \partial_x u_z \) \\ \hline
\( S_{36} \) & \( 2 \partial_z u_y + n_7 \partial_y u_z \) \\ \hline
\( n_a \) & \( n+a \) \\ \hline
\end{longtable}

The transition from 2D to 3D flow significantly increases the complexity of nonlinear constitutive relationships due to three primary factors:

(1) Increased nonequilibrium components:
    As the degrees of freedom increase, the number of viscous stress components expands from three in 2D $(xx, xy, yy)$ to six in 3D $(xx, xy, xz, yy, yz, zz)$. Similarly, the heat flux increases from two components $(x, y)$ in 2D to three $(x, y, z)$ in 3D.

(2) Greater diversity of driving forces:
    In 3D flows, additional nonequilibrium driving forces arise, including first-order derivatives such as $\partial_z \rho$, $\partial_z T$, and $\partial_\alpha u_\beta$ ($\alpha = z$ and/or $\beta = z$). Additionally, second-order derivatives and cross terms involving the $z$-direction appear, including:
    $\frac{\partial^2}{\partial x \partial z} \rho$, $\frac{\partial^2}{\partial y \partial z} \rho$, $\frac{\partial^2}{\partial z^2} \rho$, $\frac{\partial^2}{\partial x \partial z} u_x$, $\frac{\partial^2}{\partial y \partial z} u_y$, $\frac{\partial^2}{\partial x \partial z} u_z$, $\frac{\partial^2}{\partial y \partial z} u_z$, $\frac{\partial^2}{\partial x^2} u_z$, $\frac{\partial^2}{\partial y^2} u_z$, $\frac{\partial^2}{\partial z^2} u_z$.

(3) Enhanced cross-coupling among driving forces:
    The cross-coupling among nonequilibrium driving forces becomes highly intricate, involving terms such as:
    $\partial_x u_x \partial_z u_z$,
$\partial_y u_x \partial_z u_z$,
$\partial_z u_x \partial_z u_z$,
$\partial_x u_y \partial_z u_z$,
$\partial_y u_y \partial_z u_z$,
$\partial_z u_y \partial_z u_z$,
$\partial_x u_z \partial_z u_z$,
$\partial_y u_z \partial_z u_z$,
$\partial_x u_x \partial_z u_x$,
$\partial_y u_y \partial_z u_x$,
$\partial_z u_y \partial_z u_x$,
$\partial_x u_x \partial_z u_y$,
$\partial_y u_y \partial_z u_y$,
$\partial_x u_x \partial_x u_z$,
$\partial_x u_y \partial_x u_z$,
$\partial_y u_y \partial_x u_z$,
$\partial_y u_x \partial_y u_z$,
$\partial_x \rho \partial_z \rho$,
$\partial_y \rho \partial_z \rho$,
$\partial_x T \partial_z T$,
$\partial_y T \partial_z T$,
$\partial_z u_x \partial_x T$,
$\partial_x u_z \partial_x T$,
$\partial_z u_y \partial_y T$,
$\partial_y u_z \partial_y T$,
$\partial_z u_z \partial_y T$,
$\partial_x u_x \partial_z T$,
$\partial_z u_x \partial_z T$,
$\partial_y u_y \partial_z T$,
$\partial_z u_y \partial_z T$,
$\partial_x u_z \partial_z T$,
$\partial_y u_z \partial_z T$.

On the one hand, the additional terms reflect the intrinsic complexity of the system's nonequilibrium mechanisms, effects, and behaviors. This complexity intensifies competition and coordination among diverse driving forces across broader spatiotemporal scales. Consequently, the system's nonlinearity is substantially amplified.

On the other hand, these additional terms enhance and extend the model's capability to describe flow behavior and heat transfer, enabling it to capture a wider range of spatial scales and intricate physical interactions. The interplay among driving forces shifts from independent effects to intricate nonlinear interactions.
Overcoming these challenges necessitates advanced mathematical modeling and numerical methods for practical applications.

Finally, eliminating macroscopic quantities and gradients related to the $z$-direction reduces the 3D case to its 2D counterpart. This simplification highlights the link between spatial dimensions and the complexity of nonequilibrium dynamics.

%This heightened complexity is reflected in the diversity of physical phenomena and the increased nonlinearity of constitutive models and governing equations.

%As a result, the degree of nonlinearity in the system is substantially amplified. %, emphasizing the necessity for improved models.

\section*{Data Availability}
The data that support the findings of this study are available from the corresponding author upon reasonable request.

%\appendix
\section*{References}
\bibliography{3DDBM-Burnett}% Produces the bibliography via BibTeX.

\end{document}